\documentclass[aps,prd,preprintnumbers,superscriptaddress,nofootinbib,floatfix,twocolumn,notitlepage]{revtex4-2}
\usepackage{paralist}
\usepackage{amsmath}
\usepackage{geometry}

\usepackage{dcolumn}   
\usepackage{amssymb,mathrsfs}\usepackage{hyperref}
\hypersetup{
    colorlinks  = true,
    linkcolor   = [rgb]{0,0.3,0.7},
    citecolor   = [rgb]{0,0.3,0.7},
    urlcolor    = [rgb]{0,0.3,0.7},
    linktocpage = true
}
\usepackage{gensymb}
\usepackage{slashed}
\usepackage{graphicx,epsfig,dcolumn,bm,xspace}
\usepackage{booktabs}
\usepackage{supertabular}
\usepackage{multirow}
\usepackage{cleveref}
\crefname{equation}{Eq.}{Eqs.}
\crefname{figure}{Fig.}{Figs.}
\usepackage{bbold}
\usepackage{titlesec}\usepackage{enumitem}
\setlist{topsep=4pt, partopsep=0pt} 

\usepackage{xcolor}
\usepackage{nicematrix}
\usepackage{graphicx}

\begin{document}

\newcommand{\ovbb}[0]{$0\nu\beta\beta$ }
\newcommand{\lr}[1]{\left( #1\right)}
\newcommand{\lrs}[1]{\big( #1\big)}
\newcommand{\sfrac}[2]{ #1\big/ #2}
\newcommand{\hc}{\text{h.c.}}
\newcommand{\eq}{\text{eq}}
\newcommand{\im}{\text{Im}}
\newcommand{\re}{\text{Re}}
\newcommand{\tr}{\text{Tr}}
\newcommand{\BR}[1]{\text{BR}\big( #1\big)}
      
\newcommand{\ve}[1]{\ensuremath{\vec{\bf{#1}}}\xspace}
\newcommand{\p}{\ve{p}}

\newcommand{\ket}[1]{|{#1} \rangle}
\newcommand{\bra}[1]{\langle {#1}|}
\newcommand{\mean}[1]{\left\langle {#1}\right\rangle}
\newcommand{\bk}[2]{\langle{#1}|{#2}\rangle}
\newcommand{\mel}[3]{\langle{#2}|{#1}|{#3}\rangle}

\newcommand{\gmu}{\gamma^{\mu}}
\newcommand{\gnu}{\gamma^{\nu}}
\newcommand{\gsig}{\gamma^{\sigma}}
\newcommand{\grho}{\gamma^{\rho}}
\newcommand{\gMu}{\gamma_{\mu}}
\newcommand{\gNu}{\gamma_{\nu}}
\newcommand{\gSig}{\gamma_{\sigma}}
\newcommand{\gRho}{\gamma_{\rho}}
\newcommand{\gfive}{\gamma_5}

\newcommand{\uspinor}[2]{u_{#1}(\vec{#2})}
\newcommand{\buspinor}[2]{\bar{u}_{#1}(\vec{#2})}
\newcommand{\vspinor}[2]{v_{#1}(\vec{#2})}
\newcommand{\bvspinor}[2]{\bar{v}_{#1}(\vec{#2})}

\newcommand{\qrecsq}{q_{\rm rec}^2}
\newcommand{\qrec}{q_{\rm rec}}

\title{Smuggling baryon asymmetry through strong washout}

\author{K\r{a}re Fridell}
\email{karef@chalmers.se}
\affiliation{Institute of Particle and Nuclear Physics, Faculty of Mathematics and Physics, Charles University in Prague, V Holešovičkách 2, 180 00 Praha 8, Czech Republic}
\affiliation{Chalmers University of Technology, Department of Physics and Astronomy, SE-41296, Gothenburg,
Sweden}

\author{Michal Malinský}
\email{michal.malinsky@matfyz.cuni.cz}
\affiliation{Institute of Particle and Nuclear Physics, Faculty of Mathematics and Physics, Charles University in Prague, V Holešovičkách 2, 180 00 Praha 8, Czech Republic}

\author{Javier Perez-Soler}
\email{javier.perez.soler@ific.uv.es}
\affiliation{Instituto de Física Corpuscular, CSIC-Universitat de València, 46980 Paterna, Valencia, Spain}

\begin{abstract}
\noindent We propose a scheme in which a primordial net baryon number generated at or around the unification scale may survive subsequent strong washout phases associated with active sphalerons and interactions of right-handed neutrinos inherent to thermal leptogenesis. A simple realization of this, namely, a long-lived scalar population injecting the net $B-L$ it carries into the Standard Model matter sector late enough to be safe from a strong washout, is constructed and studied in the context of a simple Pati-Salam scalar sector with a heavy $SU(4)_C$ sextet scalar. 

\end{abstract}

\maketitle

\section{Introduction} \label{sec:intro}
%
The $CP$- and $B-L$\,-violating decay of an out-of-equilibrium population of heavy Majorana neutrinos assisted by $B-L$ -conserving sphaleron transitions~\cite{Kuzmin:1985mm,Davidson:2008bu} (the so called baryogenesis through leptogenesis mechanism~\cite{Fukugita:1986hr}) represents one of the most compelling theoretical scenarios in which the observed baryon asymmetry of the Universe can emerge dynamically and naturally in beyond-Standard-Model (BSM) physics. 
On its own, this scheme is not very predictive though, owing to the disparity between the number of relevant low-energy observables and the plethora of a-priori unknown parameters encountered even in its simplest realizations such as type-I seesaw~\cite{Minkowski:1977sc} with 3 right-handed neutrinos (RHNs) simply added to the Standard Model (SM). 
Moreover, as suggested by Davidson and Ibarra~\cite{Davidson:2002qv}, the natural domain of such new dynamics typically (though not necessarily) falls many orders of magnitude beyond the electroweak scale, thus further  reducing the prospects of directly probing the underlying layers of reality.

The situation becomes much better if leptogenesis is considered in schemes in which the relevant parameters happen to be subject to additional constraints. This often occurs in unified scenarios in which the lepton-sector Yukawa couplings are correlated to those governing the quarks. In such cases the extra constraint from the observed baryon-to-photon number density ratio~\cite{Planck:2018vyg}
\begin{equation}
    \eta_B^\text{obs}=(6.20\pm 0.15)\times 10^{-10}
\end{equation}
may be unexpectedly powerful (see e.g.~\cite{Babu:2024ahk} and references therein).
On the other hand, the $B$ and $L$ violation inherent to the unification-scale ($M_U$) dynamics of these models may complicate the eventual relic asymmetry calculation as it offers other potential sources of net baryon and lepton number generation, such as e.g.\ $B$-violating decays of colour-triplet scalars, c.f.\ Ref.~\cite{Langacker:1980js}. For large-enough gap between $M_U$ and the RHN masses, such "primordial" asymmetry would then represent a non-trivial initial condition for the subsequent leptogenesis stage.  

Usually, one assumes that the yield of such a unification-specific baryogenesis stage before the RHNs getting out of equilibrium is mostly irrelevant in the leptogenesis context, for at least two reasons: 
    First, the washout due to the $\Delta L\neq 0$ processes inherent to leptogenesis, assisted by sphalerons (whose effects become significant below about $T_{\rm sph.}\sim 10^{13}$~GeV~\cite{Garbrecht:2014kda}), is typically strong enough to erase the primordial component to such a degree that the simple zero initial asymmetry asumption for the subsequent leptogenesis stage becomes fully justified. Second, in popular simple unifications such as the  $SO(10)$ GUTs~\cite{Fritzsch:1974nn} or their Pati-Salam~\cite{Pati:1974yy} or left-right (LR) descendants~\cite{PhysRevD.11.566,PhysRevD.12.1502}, this is further facilitated by the fact that the the $U(1)_{B-L}$ symmetry often survives the unified symmetry breaking and $B-L$ remains conserved (and zero) well below $M_U$. In such a case, the primordial asymmetry is generated solely in the $B+L$ sector and, at the onset of leptogenesis (corresponding roughly to the $U(1)_{B-L}$ symmetry breaking scale $M_{B-L}$, typically well below $T_{\rm sph.}$, c.f.\ Ref.~\cite{Bertolini:2009qj}), such a net $B+L$ is already gone. 

However, the unification-scale baryogenesis as a precursor to the usual leptogenesis stage may have a significant impact in frameworks in which the very dynamics at or around $M_U$ already violates $B-L$ and the subsequent washout of thus generated net yield may be for some reason inhibited. 
This, of course, has been entertained in the literature before in many variants, avoiding a strong washout e.g.\ by evading the time-window of sphaleron transitions~\cite{Davidson:1996hs,AristizabalSierra:2013lyx,Huang:2016wwj}, such that successful baryogenesis can be achieved even in the absence of $B-L$-violation, or alternatively by utilizing a non-trivial flavour structure in the initial $B$- and $L$-asymmetries generated at some high scale~\cite{Dreiner:1992vm,Cline:1993vv,Cline:1993bd,Domcke:2020quw}, which can be transferred to a remaining $B-L$ asymmetry by the gradual onset of SM Yukawa interactions in the presence of $B-L$-violating washout channels.

In the current paper we shall demonstrate that a similar outcome can be achieved even in presence of active sphalerons and irrespective of the particular flavour structure, in the situation where the primordial $B-L$ is stored in a particle population that does not talk to the leptons in any significant way, stable enough to release the asymmetry into the SM matter sector only after the RHN-driven washout encountered in standard leptogenesis becomes insignificant.

More specifically, we show how a primordial asymmetry stored in the scalar sector of a simple Pati-Salam (PS)~\cite{Pati:1974yy} model can be protected from strong washout and active sphalerons and later released into the matter sector of the SM, even in case of diagonal Yukawa couplings of the decaying heavy scalars.
To this end, let us note that heavy scalars and their decays have been associated with baryogenesis in many different contexts such as GUTs~\cite{Enomoto:2011py,Gu:2011pf,Babu:2012vb}, type-II seesaw~\cite{Hambye:2003ka,Hambye:2005tk,Fong:2025aya}, leptoquarks~\cite{Hati:2018cqp,Fridell:2024otl}, or models leading to neutron--antineutron oscillation~\cite{Babu:2006xc,Babu:2012vc,Herrmann:2014fha,Allahverdi:2017edd,Fridell:2021gag}. In these, they typically  couple directly to fermions in $CP$-violating decays, generating the baryon asymmetry in a one-step process. However, extending such scenarios by additional $B-L$-violating interactions, e.g.\ in order to accommodate $\Delta L=2$ Majorana neutrino masses (unless already implicit in the model), could potentially add new washout channels, c.f.\ e.g.\ Ref.~\cite{Buchmuller:2004nz}; strong washout may then remove the baryon asymmetry after its initial generation, thereby rendering the baryogenesis mechanism ultimately unsuccessful. 

In a wider perspective, our study also provides a counter-example to previous works in which the presence of strong washout at intermediate scales was assumed to directly impose constraints on high-scale baryogenesis, e.g.\ in the context of neutrinoless double beta decay~\cite{Deppisch:2015yqa,Deppisch:2017ecm}, rare meson decays~\cite{Deppisch:2020oyx,Endo:2026qof}, collider signatures of lepton number violation~\cite{Frere:2008ct,Deppisch:2013jxa}, or probes of $\Delta B\neq \Delta L$-interactions in general~\cite{Nelson:1990ir,Campbell:1990fa}. 
Note however that the existence of the ``primordial $B-L$ asymmetry protection mechanism'' proposed in the current work in the specific Pati-Salam context is not a feature  generally available in BSM theories. Besides having to fulfil the Sakharov conditions within a sector associated to the high-scale symmetry breaking (such as a suitable set of heavy scalars in a unified model) one needs a long-lived population capable of smuggling thus obtained net $B-L$ through later stages. This may be difficult to attain in general, but could also arise as an unintended consequence in scenarios where the longest-living scalar is sufficiently light.

The study is organized as follows: In Sec.~\ref{sec:mechanism} we discuss the core of the proposed mechanism in the context of a SM extension featuring two types of scalar fields that conspire in a specific way to protect the primordial net $B-L$ from strong washout. In Sec.~\ref{sec:patisalam} we provide a minimal realisation of the mechanism in the Pati-Salam context, calculating the spectra and couplings of the critical fields along with the relevant interaction and decay rates governing the associated Boltzmann equations. Additional constraints from proton decay and neutrino mass generation are identified in Sec.~\ref{sec:pheno}. In Sec.~\ref{sec:results} the results of a dedicated numerical analysis are presented. 
Technical details of the analysis are deferred into a set of Appendices.  

\section{The mechanism} \label{sec:mechanism}

In order to demonstrate the proposed mechanism in general terms we first consider a simple extension of the SM featuring two non-singlet high-scale scalar fields $\phi_{1,2}$, out of which the heavier field $\phi_1$ consists of two copies (``generations'') and has tree-level couplings to SM fermions, while the lighter field $\phi_2$ interacts feebly with the SM. For this scenario to dynamically generate the observed matter--antimatter asymmetry of the Universe it needs to fulfil the three Sakharov conditions~\cite{Sakharov:1967dj}.

As for the first condition ($B$-violation), it is already met by the $\Delta (B+L)=3$ sphaleron interactions within the SM. For a baryogenesis mechanism active during rapid sphaleron transitions, the first condition therefore translates into the need for $B-L$-violation. In the present simplified scenario, meeting this condition hinges on the $B-L$-charges of $\phi_1$ and $\phi_2$.
Next, the interference between tree-level and one-loop (self-energy) contributions to the decay amplitudes of $\phi_1$, via mixing with its second ``generation'', could lead to $CP$-violating decays of the $\phi\to \text{SM}\, \text{SM}$ and $\phi_1\to \phi_2\,\phi_2$ types (or alternatively $\phi_1\to\phi_2\, H^{(*)}$), thereby meeting condition 2 ($C$ and $CP$ violation). 
If, on top of that, $\phi_1$ decays out of equilibrium, Sakharov's third condition is also met. 

This generally leads to a coexistence of a $B-L$ asymmetry in the SM matter sector and an asymmetry between the $\phi_2$ and $\phi_2^*$ populations. Note, however, that both decay modes of $\phi_1$ provide equal amounts of $CP$ violation\footnote{The $CPT$ symmetry requires the total widths of $\phi_1$ and $\phi_1^*$ to be the same, such that both decay modes of $\phi_1$ must yield an equal amount of $CP$ violation.} and, hence, thus generated matter sector $B-L$- and $\phi_2$--$\phi_2^*$ asymmetries are, so far, also equal. Therefore, without further processing, a cancellation between the $\phi_2$--$\phi_2^*$ and the matter-sector $B-L$ asymmetries is expected once $\phi_2$ decays, leaving the low-energy Universe without any baryon asymmetry.

However, such a cancellation could potentially be avoided, due to a twist that we have not seen discussed in previous works. Should the decaying scalar have different branching ratios into the two modes, then washout via inverse decays will affect the two sectors differently. Therefore, the degeneracy between the asymmetries stored in $B-L$ and $\phi_2$--$\phi_2^*$ can be broken by inverse decays, preventing their later complete cancellation. Strong washout channels that affect one sector but not the other, like for instance the $LH\leftrightarrow \bar{L}H^*$  $\Delta L=2$ processes mediated by Majorana RHNs, would further break the degeneracy and, hence, efficiently avoid the complete cancellation among the two sectors. 

Technically, this of course requires the absence of any strong coupling between $\phi_2$ and the SM so that the $\phi_2$--$\phi_2^*$ asymmetry survives through the other sector's strong washout phase, in order to decay only at later stages transferring the asymmetry it carries into the SM when it is already ``safe'' to do so. This, in turn, may be viewed as another defining feature of potentially successful settings of this kind; as we shall see, the main virtue of the Pati-Salam scenario elaborated on in Sec.~\ref{sec:patisalam} is that it implements this prerequisite in a very natural way.

\subsection{Simple analytic $\eta_B$ estimate}
Assuming a complete washout of the matter-sector $B-L$ at intermediate scales, we estimate the final baryon asymmetry (due to the delayed $\phi_2$ decay) as
\begin{equation}
    \eta_B\approx \frac{1}{79}\epsilon\,\eta_{\phi_1^+}^0\, .
\end{equation}
Here $\epsilon$ is the amount of $CP$-violation in $\phi_1$-decays, $\eta_{\phi_1^+}^0$ is the initial average abundance of $\phi_1$ and $\phi_1^*$, such that $\eta_{\phi_1^+}^0\equiv \frac{1}{2}\big(\eta_{\phi_1}^0+\eta_{\phi_1^*}^0\big)$, and the factor $1/79$ is a combination of the sphaleron conversion factor $28/79$~\cite{Giudice:2003jh} and a factor $g_*(T_*)/g_*(T_0)\approx 1/28$ accounting for the change in relativistic degrees of freedom between the early universe (at temperature $T_*$) and the present scale $T_0$. 

Alternatively, for an incomplete intermediate-scale washout of the matter sector $B-L$ we can estimate the final baryon asymmetry from the relevant system of Boltzmann equations
\begin{align}
        \frac{d\eta_{\phi_2^-}}{dz} &= -D_{\phi_2}\eta_{\phi_2^-}\,,\\
        \frac{d\eta_{B-L}}{dz} &= - W_{N}\eta_{B-L}-D_{\phi_2}\eta_{\phi_2^-}\,,
\end{align}
where $\eta_{\phi_2^-}\equiv \frac{1}{2}\lr{\eta_{\phi_2}-\eta_{\phi_2^*}}$ and $z\equiv m_{\phi_2}/T$.
The washout function $W_{N}$ can be written as
\begin{equation}
    W_{N}(z)=\frac{z}{6\zeta(3)}\frac{m_N}{m_{\phi_2}}\frac{\Gamma_N}{H}K_1\Big(\frac{m_N}{m_{\phi_2}}z\Big)\, ,
\end{equation}
where $H$ is the Hubble rate; note that for the sake of simplicity we have kept only the on-shell $s$-channel part of $LH\leftrightarrow \bar{L}H^*$ via $N$ (see Appendix~\ref{app:boltzmann} for details). For far-out-of-equilibrium decays of $\phi_2$, one can approximate the decay function $D_{\phi_2}$ as
\begin{equation}
\begin{aligned}
    D_{\phi_2}(z) &\approx \frac{\Gamma_{\phi_2}}{H}\frac{1}{z}\, .
    \end{aligned}
\end{equation}
Note that $H\propto z^{-2}$ such that $D_{\phi_2}\propto z$. Finally, the relevant boundary conditions for an initial $z_i\ll 1$ read
\begin{equation}
    \eta_{B-L}\Big|_{z=z_i} = \eta_{\phi_2^-}\Big|_{z=z_i} \approx -\epsilon\,\eta_{\phi_1^+}^0\, .
\end{equation}
Assuming that $W_N$ peaks at some $z$ for which $D_{\phi_2}$ is still small, such that the two regimes can be separated and solved independently, we find that, in the asymptotic $z\gg 1$ limit the final baryon asymmetry can be expressed as
\begin{equation}
\label{eq:etaBapproximation}
    \eta_{B} \approx \frac{1}{79}\epsilon\, \eta_{\phi_1^+}^0(1-e^{-\mathcal{W}_N^\infty})\, ,
\end{equation}
with
\begin{equation}
    \mathcal{W}_N^\infty\equiv\int_{z_i}^\infty W_N(z')dz'\,.
\end{equation}
We see that a stronger intermediate-scale washout of $B-L$ in the matter sector indeed leads to a larger final baryon asymmetry.

\section{A Pati-Salam model with heavy scalar sextet} \label{sec:patisalam}
%
\subsection{The model}\label{sec:model}
\subsubsection{Field content}
As an explicit and near-minimal realisation of the mechanism described in Sec.~\ref{sec:mechanism} we  consider a model based on the the Pati-Salam symmetry group~\cite{Pati:1974yy} $G_\text{PS}=SU(4)_C\times SU(2)_L \times SU(2)_R$ broken directly to the $G_\text{SM}=SU(3)_c\times SU(2)_L \times U(1)_Y$ by the SM-singlet VEV of the $\Phi_R\equiv(\mathbf{\overline{4}},\mathbf{1},\mathbf{2})$ scalar multiplet, accompanied by an $SU(4)_C$ sextet $\Omega\equiv(\mathbf{6},\mathbf{1},\mathbf{1})$. At the Pati-Salam breaking scale their colour-triplet components mix, giving rise to a pair of massive scalars $\Delta_1$ and $\Delta_2$, that will play the role of the $\phi_1$-type system\footnote{Note that a sextet of $SU(4)_C$ can have tree-level Yukawa couplings (forbidden for the quadruplet), leading to enhanced $CP$-violation in decays that involve fermionic self-energy loops.} of Sec.~\ref{sec:mechanism}. The lighter, long-lived $\phi_2$ field therein will then correspond to the colour-triplet $SU(2)_L$-doublet component $\Upsilon$ of $\Phi_L\equiv (\mathbf{4},\mathbf{2},\mathbf{1})$, the $L$--$R$ partner of $\Phi_R$. The complete field structure of the model, including the minimal set of fermions accommodating the SM matter multiplets is given in Table~\ref{tab:fields}.  
\renewcommand{\arraystretch}{1.25}
\begin{table}
    \centering
    \begin{tabular}{l l | l l r}
    PS field & $G_\text{PS}$ & SM field &$G_\text{SM}$ & $B-L$\\
    \hline
    Matter\\
    \hline
    $\Psi_L$ & $(\mathbf{4},\mathbf{2},\mathbf{1})$ & $Q$ & $(\mathbf{3},\mathbf{2},+1/6)$ & $+1/3$ \\
    & & $L$ & $(\mathbf{1},\mathbf{2},-1/2)$ & $-1$ \\[1mm]
    $\Psi_R$ 
    & $(\overline{\mathbf{4}},\mathbf{1},\mathbf{2})$ & $u^c$ & $(\overline{\mathbf{3}},\mathbf{1},-2/3)$ & $-1/3$ \\
    && $d^c$ & $(\overline{\mathbf{3}},\mathbf{1},+1/3)$  & $-1/3$ \\
    &  & $e^c$ & $(\mathbf{1},\mathbf{1},+1)$ & $+1$\\
    && $N^c$ & $(\mathbf{1},\mathbf{1},0)$ & $+1$ \\
    \hline
    Gauge\\
    \hline
      $V_\mu$  & $(\mathbf{15},\mathbf{1},\mathbf{1})$ & $X_\mu$ &$(\mathbf{3},\mathbf{1},+2/3)$ & $+4/3$ \\
      &&$X_\mu^\dagger$&$(\overline{\mathbf{3}},\mathbf{1},-2/3)$ &  $-4/3$\\
      &&$G_\mu$&$(\mathbf{8},\mathbf{1},0)$ & $0$\\
       &&$B'_\mu$& $(\mathbf{1},\mathbf{1},0)$ & $0$\\[1mm]
      $W_L{}_\mu$&  $(\mathbf{1},\mathbf{3},\mathbf{1})$&$W_\mu$&$(\mathbf{1},\mathbf{3},0)$ & $0$\\[1mm]
      $W_R{}_\mu$&  $(\mathbf{1},\mathbf{1},\mathbf{3})$&$W'_\mu$&$(\mathbf{1},\mathbf{1},+1)$ & $0$\\
      &  &$W'_\mu{}^\dagger$&$(\mathbf{1},\mathbf{1},-1)$ & $0$\\
      &  &$Z'_\mu$&$(\mathbf{1},\mathbf{1},0)$ & $0$\\
    \hline
    Scalar\\
    \hline
       $\Sigma$ & $(\mathbf{1},\mathbf{2},\mathbf{2})$ & $H$ & $(\mathbf{1},\mathbf{2},+1/2)$ & $0$\\
       && $\Tilde{H}$ & $(\mathbf{1},\mathbf{2},-1/2)$ & $0$ \\[1mm]
       $\Omega$ & $(\mathbf{6},\mathbf{1},\mathbf{1})$ & $T$ & $(\mathbf{3},\mathbf{1},-1/3)$ & $-2/3$\\
       && $T^\dagger$ & $(\overline{\mathbf{3}},\mathbf{1},+1/3)$ &  $+2/3$ \\[1mm]
    $\Phi_L$ & $(\mathbf{4},\mathbf{2},\mathbf{1})$ & $\Upsilon$ & $(\mathbf{3},\mathbf{2},+1/6)$ & $+1/3$ \\
    & & $\tilde{H'}$ & $(\mathbf{1},\mathbf{2},-1/2)$ & $-1$ \\[1mm]
      $\Phi_R$ & $(\overline{\mathbf{4}},\mathbf{1},\mathbf{2})$ &   $T'{}^\dagger$  & $(\overline{\mathbf{3}},\mathbf{1},+1/3)$ & $-1/3$\\
    && $S$ & $(\mathbf{1},\mathbf{1},0)$ & $1$ \\
    \hline
    \end{tabular}
    \caption{Field content of the model under consideration with matter, gauge and scalar fields' transformation properties under the Pati-Salam symmetry including the  SM multiplets within and their $B-L$ charges. Note that, due to the pseudoreality of $\mathbf{2}$ of $SU(2)_L$ and $\mathbf{6}$ of $SU(4)_C$, what we denote as $H$ and $\tilde{H}$ are not independent fields (indeed, they are connected as usual via $\tilde{H}=i\sigma_2 H^*$), and neither are $T$ and $T^\dagger$ (as $SU(3)_c$ components of the $SU(4)_C$ sextet). The benefit of the current notation is that later on, cf.\ Sec.~\ref{sec:spectrum}, the fields that shall mix will be $H$ with $H'$ and $T$ with $T'$ (with no need for further conjugation anywhere).}
    \label{tab:fields}
\end{table}

\subsubsection{Interactions}\label{sec:lagrangian}
The gauge sector of the model is fully defined by the covariant derivative 
\begin{equation}
    D_\mu=\partial_\mu-ig_4T_4^aV^a_\mu-ig_LT_L^iW_L{}^i_\mu-ig_RT_R^jW_R{}^j_\mu\;,
\end{equation}
where $T_4^a$ and $\frac{1}{2}\sigma^a$ are the generators of the $SU(4)_C$ and $SU(2)$ gauge factors, respectively, and the corresponding gauge fields are given in Tab.~\ref{tab:fields}.

The Yukawa couplings of the fermionic fourplets $\Psi_L$ and $\Psi_R$ with the $\Sigma$ ($L$--$R$ bidoublet) and $\Omega$ ($SU(4)_C$-sextet) scalars (see again Tab.~\ref{tab:fields}) is given by
\begin{equation}
\begin{aligned}
\label{eq:Yukawa}
\mathcal{L}_Y &= Y_{LR} \overline{\Psi_L^c}\Psi_R\Sigma^* + Y_{R}\epsilon_{ab}\overline{\Psi_R^c}{}^{a}\Psi_R{}^{b}\Omega+\hc
\end{aligned}
\end{equation}
Note that the setting with only one copy of the $L$--$R$ bidoublet employed here does not provide enough room to reproduce all the SM fermion masses and mixing parameters, and a renormalizable model with a fully realistic low-energy limit would indeed require more such fields. This, however, is not the scope of the current study, and the presence of additional bidoublets would change nothing on the argument we wish to make. Hence, for the sake of simplicity we shall stick to such a minimal setting here.    

Actually, a much more important feature of the Yukawa sector is the absence of tree-level couplings of $\Phi_L$ and $\Phi_R$ to matter (and, in particular $\Upsilon$ within) which are forbidden at the high scale by the gauge symmetry. 
Such interactions naturally emerge only at the level of $d=5$ operators 
\begin{equation}
\label{eq:ops12}
    \begin{aligned}
    \mathcal{O}_{LR2} &= \overline{\Psi}_L{}_{\alpha i} \Psi_R^c{}^\alpha_{\;\, a}\Phi_L{}^{\beta i} \Phi_R{}_\beta^{\;\,a}\, ,\\
    \mathcal{O}_{RR2} &= \overline{\Psi}_R{}^{\alpha}_{\;\,a} \Psi_R^c{}^{\beta}_{\;\,b}\Phi_R{}_{\alpha}^{\;\,a}\Phi_R{}_{\beta}^{\;\,b}\, ,
    \end{aligned}
\end{equation}
and their hermitenan conjugates.

Finally, the scalar potential that we shall consider in this section can be symbolically (i.e., with all indices dropped) written as
\begin{equation}\label{eq:IIIA_scalarPotential}
\begin{aligned}
    V \ni
    \,&m_L^2 \Phi_L\Phi_L^*\! + m_R^2\Phi_R\Phi_R^*\! + \frac{m_\Sigma^2}{2}\Sigma\Sigma^*\! +\frac{m_\Omega^2}{4}\Omega\Omega^*\\ 
    +\,&\Big[\frac{\mu_R}{2}\epsilon_{ab}\Phi_R{}^{a}\Phi_R{}^{b} \Omega + \mu_{LR}\Phi_L\Phi_R\Sigma^*+\hc\Big]\\
    +\,&\lambda_5\lr{\Phi_R\Phi_R^*}^2+\lambda_6\tr\Big[\Phi_R\Phi_R^*\Phi_R\Phi_R^*\Big]\, ,
\end{aligned}
\end{equation}
where $\mu_R$ and $\mu_{LR}$ are (in principle complex) parameters with mass dimension 1, and $\lambda_5$ and $\lambda_6$ are dimensionless real couplings. 

Note that \cref{eq:Yukawa,eq:ops12,eq:IIIA_scalarPotential} do not
represent the full set of operators one can contemplate at the given level --- for simplicity we have omitted all pieces that are neither essential nor potentially disturbing for the baryogenesis mechanism under consideration (c.f.\ Sec.~\ref{sec:mechanism}).
For a more complete account of the $d=4$ and $d=5$ structure of the model, including full expansion of the contractions in~\cref{eq:Yukawa,eq:IIIA_scalarPotential}, the reader is deferred to Appendix~\ref{app:fullmodel}.

\subsubsection{The Pati-Salam symmetry breaking and matching}
The $SU(4)_C\times SU(2)_R$ part of the Pati-Salam symmetry is broken to the $SU(3)_c\times U(1)_Y$ part of the SM by a vacuum expectation value (VEV) of the SM singlet $S$ of $\Phi_R$ (c.f.\ Appendix~\ref{app:patisalam}) 
\begin{equation}
    \langle\Phi_R^{42}\rangle=V_G\, .
\end{equation}
In the standard convention with the SM hypercharge embedded into the Pati-Salam group as 
\begin{equation}
Y=T_R^3+\sqrt{\tfrac{2}{3}}T_{15}\,,    
\end{equation}
the low-energy $g_c$ and $g_Y$ couplings are matched (at $V_G$) to the high-energy ones as
\begin{equation}
\label{eq:gYgc}
    g_c=g_4\,,\qquad g_Y=\frac{g_4g_R}{\sqrt{g_4^2+\frac{2}{3}g_R^2}}\, .
\end{equation}

The natural requirement of stability of this vacuum leads to 
\begin{equation}
    m_R^2=-2\lr{\lambda_5+\lambda_6}V_G^2\, ,
\end{equation}
along with
\begin{equation}
\label{eq:constraint4}
    \lambda_6+\lambda_5 > 0 > \frac{|\mu_R|^2}{2m_{\Omega}^2} +\lambda_5\,,
\end{equation}
where the pair of inequalities in~\cref{eq:constraint4} ensures the absence of tachyons (in the $T$--$T'$  and $S$ sectors).

\subsubsection{The spectrum\label{sec:spectrum}}
The tree-level heavy vector boson masses are calculated readily
\begin{equation}
\begin{aligned}
    & m_{W'}^2 = \frac{1}{4}V_G^2\,g_R^2,\quad m_X^2 = \frac{1}{4}V_G^2\,g_4^2,\\
    & m_{Z'}^2 = \frac{1}{4}V_G^2\left(g_R^2+\frac{3}{2}\,g_4^2\right)\, .
\end{aligned} 
\end{equation}
As for the scalars, the situation is more involved due to significant mixing.
Both $\Sigma$ and $\Phi_L$ contain fields in the $(\mathbf{1},\mathbf{2},+1/2)$ representation of the SM ($H$ and $H'$, respectively), which mix and contribute to the masses of the SM fermions upon electroweak symmetry breaking (EWSB). Similarly, $\Omega$ and $\Phi_R$ both contain scalar leptoquarks in the $(\mathbf{3},\mathbf{1},-1/3)$ representation ($T$ and $T'$, respectively), which also mix. The leptoquark $\Upsilon$ coming from $\Phi_L$ does not mix with anything, and the remaining real scalar $S$ is a radial mode around the minimum of the broken potential.

In the broken phase, the relevant part of the scalar potential corresponding to~\cref{eq:IIIA_scalarPotential} reads
\begin{align}
    V &\ni m_\Sigma^2HH^* + m_L^2 H'H'^*+ m_\Omega^2TT^*+ m_L^2\Upsilon\Upsilon^*\nonumber\\
    & +4\lr{\lambda_5+\lambda_6}V_G^2S^2-2\lambda_6V_G^2T'^*T'\nonumber\\
    \label{eq:brokenV}
    &+\Big[\mu_RV_GTT'^*+\hc\Big]-\mu_{LR}V_GHH'^*\\
    &+ 4\lambda_5V_GST'T'^*+4\lr{\lambda_5+\lambda_6}V_GS^3\nonumber\\
    &+ \Big[\mu_{R}STT'^*-\mu_{LR}SHH'^*+\mu_{LR}T'^*\Upsilon H^*+\hc\Big]\nonumber\\
    &+\lr{\lambda_5+\lambda_6}\lr{T'T'^*T'T'^*+4S^4}+4\lambda_5S^2T'T'^*\, .\nonumber
\end{align}
In what follows, we denote the corresponding mass eigenstates of $T$--$T'$ and $H$--$H'$ as $\Delta_{1,2}$ and $H_{1,2}$, respectively. These are given by
\begin{align}\label{eq:UDeltaUHmatrices}
    \begin{pmatrix}
        \Delta_1\\
        \Delta_2
    \end{pmatrix} = U_\Delta \begin{pmatrix}
        T\\
        T'
    \end{pmatrix}\, ,\quad    
    \begin{pmatrix}
        H_1\\
        H_2
    \end{pmatrix} = U_H \begin{pmatrix}
        H\\
        H'
    \end{pmatrix}\, .
\end{align}
The corresponding mixing matrix elements are
\begin{align}
        &U_\Delta{}_{ij} = \frac{\delta_{j1}\lrs{m_{\Delta_i}^2+2\lambda_5V_G^2}+\delta_{j2}\mu_RV_G}{\sqrt{|\mu_R|^2V_G^2+\lrs{m_{\Delta_i}^2+2\lambda_5V_G^2}^2}}\, ,\\
        &U_H{}_{ij} = \frac{\delta_{j1}(m_{H_i}^2-m_L^2)-\delta_{j2}\mu_{LR}V_G}{\sqrt{|\mu_{LR}|^2V_G^2+(m_{H_i}^2-m_L^2)^2}}\, ,
\end{align}
with the masses
\begin{align}
&\begin{aligned}
    m_{\Delta_{1,2}}^2 &=\frac{1}{2}\Big(m_{\Omega}^2-2\lambda_5V_G^2\\
    &\qquad\mp \sqrt{\lr{m_{\Omega}^2+2\lambda_5V_G^2}^2+4|\mu_R|^2V_G^2}\Big)\, ,
\end{aligned}\\
&\begin{aligned}
    m_{H_{1,2}}^2 &=\frac{1}{2}\Big(m_{\Sigma}^2+m_{L}^2\\
    &\qquad\mp \sqrt{(m_{\Sigma}^2-m_{L}^2)^2+4|\mu_{LR}|^2V_G^2}\Big)\, .
\end{aligned}
\end{align} 
Finally, the masses of the remaining scalar fields are 
\begin{equation}
    m_{\Upsilon}^2 = m_L^2,\quad m_S^2 = 4\lr{\lambda_5+\lambda_6}V_G^2\, .
\end{equation}
Similarly, the broken-phase Yukawa Lagrangian is given by
\begin{equation}
\begin{aligned}
    \mathcal{L}_{Y} &= 
    Y_{LR}H^*Q\bar{d} + Y_{LR}H^i\epsilon_{ij}Q^j\bar{u} +Y_{LR}H^*L\bar{e}\\
    &+Y_{LR}H^i\epsilon_{ij}L^j\bar{N}+2Y_R\epsilon_{xyz}T^x\bar{d^c}{}^yu^z\\
    & +2Y_R T^*\bar{d^c}N  -2Y_R T^*\bar{u^c}e +\hc \, ,
\end{aligned}
\end{equation}
where $i$, $j$ and $x$, $y$, $z$ are $SU(2)_L$ and $SU(3)_c$ indices, respectively. 

\begin{figure}
    \centering
    \includegraphics[scale=1.3]{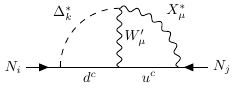}
    \caption{Example of a two-loop radiative RHN  Majorana mass generated in the Pati-Salam model under consideration. The complete set of relevant diagrams can be found in  Appendix~\ref{app:fullmodel}.}
    \label{fig:twoloopMN}
\end{figure}

The Yukawa couplings of $T$, together with the gauge interactions and the $T$--$T'$ mixing, lead to a two-loop Majorana mass for right-handed neutrinos~\cite{Witten:1979nr,ArbelaezRodriguez:2013kxw}, as shown in Fig.~\ref{fig:twoloopMN}. Following Ref.~\cite{Harries:2018tld}, we estimate the leading loop contribution as
\begin{equation}
\label{eq:2loopMN}
    M_N \approx \sum_i\frac{6g_4^2g_R^2}{(4\pi)^4}|U_{\Delta}{}_{1i}U_{\Delta}^*{}_{2i}|V_GY_R\, .
\end{equation}
Note that, for the sake of simplicity, we have neglected the $X$--$W'$ mass difference, as well as some other $\mathcal{O}$(1) factors, with essentially no impact on final results. Finally, the effective operators~\eqref{eq:ops12} yield in the broken phase
\begin{equation}
\begin{aligned}
    \mathcal{L}_\text{eff} &\supset C_{LR2}V_G\,\bar{L}N\Tilde{H'}+C_{LR2}V_G^2\, \bar{Q}u\tilde H'\\
    &C_{RR2}V_G^2\, \bar{N^c}N+ C_{RR2}V_G\bar{d^c}NT +\hc \, ,
\end{aligned}
\end{equation}
where only the $V_G$-proportional terms have been kept. The first, second and fourth terms provide direct couplings between the SM fermions and scalars, while the third term leads to a tree-level $d=5$ Majorana mass for RHNs, namely
\begin{equation}
\label{eq:baremass}
    M_N^{\rm eff} = C_{RR2}V_G^2\, .
\end{equation}
This can be almost arbitrarily large, provided that the underlying UV completion remains perturbative and the corresponding scale can be integrated out. The loop contribution of Eq.~\eqref{eq:2loopMN} can then be viewed as a conservative lower limit on the RH neutrino mass scale.
\subsection{Basic phenomenological constraints\label{sec:pheno}}
Prior to coming to the details of baryogenesis in the current scenario, let us take a look at the main phenomenological constraints it is subject to, namely, proton lifetime and neutrino mass generation via type-I seesaw. We note that, depending on the model details, additional phenomena such as e.g.\ gravitational waves, 
neutrinoless double beta decay and perhaps even collider signatures may be generated. However, in this work, we shall not pursue these directions any further. 
\subsubsection{Proton decay}
The heavy scalar triplets $\Delta_{1,2}$ of Sec.~\ref{sec:patisalam}
have both leptoquark- and diquark couplings (to the $\overline{e^c}u$ and $\overline{d}u^c$ bilinears, respectively); thus, they can mediate tree-level proton decay modes such as $p\to\pi^0e^+$, corresponding to the effective operator
\begin{align}
    \mathcal{O}_{eu^2d}^{mnk\ell}=\epsilon_{xyz}\lr{\overline{e^c}{}^mu^x{}^n}\lr{\overline{u^c}{}^y{}^kd^z{}^\ell}\, ,
\end{align}
with which we associate the Pati-Salam-scale Wilson coefficient
\begin{equation}
\label{eq:protondecaywilson}
    C_{eu^2d}^{mnk\ell}=\sum_i|U_{\Delta}{}_{1i}|^2\frac{4Y_R{}^{mn}Y_R^\dagger{}^{k\ell}}{m_{\Delta_i}^2}\, ,
\end{equation}
where $m$, $n$, $k$, and $\ell$ are flavor indices. 
Neglecting the electron mass, we find the decay width
\begin{equation}
    \Gamma(p\to e^+\pi^0)=\frac{m_p}{32\pi}\bigg[1-\frac{m_{\pi^0}^2}{m_p^2}\bigg]^2\big|C_{eu^2d}^{1111}W^R_{e^+}\big|^2\, ,
\end{equation}
where $W^R_{e^+}\approx 0.1\text{ GeV}^2$ is a lattice constant~\cite{Yoo:2021gql}. To evolve $C_{eu^2d}^{mnk\ell}$ of Eq.~\eqref{eq:protondecaywilson} to the $m_Z$ scale we use~\cite{Alonso:2014zka}
\begin{equation}
    (4\pi)^2\mu\frac{d}{d\mu}C_{eu^2d}^{mnk\ell}=(-4g_3^2-2g_1^2)C_{eu^2d}^{mnk\ell}\, ,
\end{equation}
where $\beta_1=\frac{41}{10}$ and $\beta_3=-7$ are the usual SM beta-coefficients for the $g_1=\sqrt{5/3}\,g_Y$ and $g_3$ running (c.f.\ Ref.~\cite{Mihaila:2012pz}). We also include an additional factor $A_L=1.4$ to match between the $m_Z$- and proton decay scales~\cite{Nath:2006ut}.

The most stringent limit on proton decay in the $p\to\pi^0e^+$ channel comes from the Super-Kamiokande experiment~\cite{Super-Kamiokande:2020wjk}
\begin{equation}
    \tau_{p\to \pi^0e^+}^\text{SK} > 2.4\times 10^{34}\text{ years, 90\% C.L.}
\end{equation}
In the future, this limit will be surpassed by Hyper-Kamiokande~\cite{Hyper-Kamiokande:2018ofw}, which after a 1.9 megaton-year exposure will be sensitive to 
\begin{equation}
    \tau_{p\to \pi^0e^+}^\text{HK} \lesssim 7.8\times 10^{34}\text{ years, 90\% C.L.}
\end{equation}
In what follows, we shall use these limits to put constraints on the Pati-Salam breaking scale and Yukawa couplings of the heavy triplets.
\subsubsection{Neutrino masses}\label{sec:phenomnu}
The high-scale RHN Majorana mass $M_N$, as realised in the current model (see~\cref{eq:2loopMN} of Sec.~\ref{sec:spectrum}), generates a mass matrix for the light (predominantly left-handed) neutrinos via the type-I seesaw formula
\begin{equation}
    M_\nu = v^2\,Y_{LR}M_N^{-1}Y_{LR}^T \, ,
\end{equation}
where $v$ is the SM Higgs VEV and $Y_{LR}$ is the relevant Yukawa structure. We emphasize that, in what follows, we shall use this formula mainly as a tool to estimate the RHN mass scale rather than as an accurate account of all nuances of the leptonic spectra and mixing.

Laboratory-based constraints on the light neutrino mass scale often refer to the sum of the electron row of $M_\nu$, namely
\begin{equation}
\label{eq:numasssum}
    m_\nu\equiv \sqrt{\sum_i U_{ei}^2m_i^2}\, ,
\end{equation}
where $U$ is the PMNS matrix. Using the central values of the normal ordering mixing angles and mass splittings from NuFIT 6.1~\cite{Esteban:2024eli}, 
\begin{equation}
\begin{aligned}
    &\theta_{12}=33.76^\circ,\, \theta_{23}=43.29^\circ,\, \theta_{23}=8.62^\circ, \\
    &\Delta m_{21}^2=7.54\times 10^{-5}\text{ eV}^2,\\
    &\Delta m_{31}^2=2.51\times 10^{-3}\text{ eV}^2\, ,
\end{aligned}
\end{equation}
one finds 
\begin{equation}
\label{eq:mnurange}
    8.9\times 10^{-3}\text{ eV} < m_\nu < 0.45\text{ eV}\, .
\end{equation}
Here the lower limit corresponds to zero lightest neutrino mass ($m_1$) while the upper (90\% C.L.) limit comes from the KATRIN experiment~\cite{KATRIN:2024cdt}.

\section{Baryogenesis in the PS sextet model}\label{sec:baryogenesis}
Finally, let us discuss baryogenesis in the current model. Recall that its ultimate seed is the $U(1)_{B-L}$ breaking VEV $V_G$ which drives the ``primordial'' $B-L$ (and $CP$) violating decays of coloured triplet scalars at the Pati-Salam scale. 
\subsection{Phase 1: $\Delta_1$ decays}
\begin{figure}[t]
    \centering
    \raisebox{-0.5\height}{
    \includegraphics[scale=1.25]{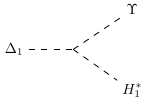}}\hspace{-1mm} 
    \raisebox{-0.5\height}{
    \includegraphics[scale=1.25]{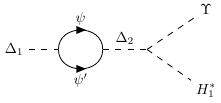}} 
    \caption{Tree-level and 1-loop decays of $\Delta_1$ into $\Upsilon$ and $H_1^*$ (left and right, respectively). Here ($\psi$, $\psi'$) represents any of the fermion pairs ($e$, $u$), ($N$, $d$) or ($u^c$, $d^c$).}
    \label{fig:CPDelta}
\end{figure}

For $m_{\Delta_1}>m_{\Upsilon}+m_{H_1}$, the second Sakharov condition is met by the interference between tree-level and 1-loop amplitudes of $\Delta_1$ decays, as shown in Fig.~\ref{fig:CPDelta}. These are, in general, rapid around $T\sim m_{\Delta_1}$, at which $\Delta_1$ becomes non-relativistic, and may go out of equilibrium, thereby meeting also the third condition. 

\begin{figure}[t]
    \centering
    \raisebox{-0.5\height}{
    \includegraphics[scale=1.25]{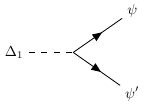}}\hspace{-1mm} 
    \raisebox{-0.5\height}{
    \includegraphics[scale=1.25]{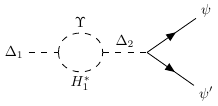}} 
    \caption{Tree-level and 1-loop decays of $\Delta_1$ into fermions (left and right, respectively).}
    \label{fig:CPDeltafermi}
\end{figure}

The $CP$ asymmetry of $\Delta_1$ decays can be readily estimated from the Feynman diagrams in Fig.~\ref{fig:CPDelta},
\begin{equation}\label{eq:delta1asymmetry}
        \epsilon = \frac{1}{\pi}U_{\Delta}{}_{11}U_{\Delta}{}_{21}^*\tr\Big[16Y_R^\dagger Y_R\Big]\im\Bigg[\frac{\mu_R^*}{\mu_R}\Bigg]\frac{x}{1-x}r\, .
\end{equation}
where 
\begin{align}
    2\epsilon & \equiv  \BR{\Delta_1\to\Upsilon H_1^*}-\BR{\Delta_1^*\to\Upsilon^* H_1}\,,\nonumber\\
    2r & \equiv  \BR{\Delta_1\to\Upsilon H_1^*}+\BR{\Delta_1^*\to\Upsilon^* H_1}\,,\nonumber
\end{align}
and $x\equiv m_{\Delta_1}^2/m_{\Delta_2}^2$ (we, of course, assume that $x$ is substantially different from 1).
A comment on the $CP$ structure of formula~\cref{eq:delta1asymmetry} is worth here: i) as expected, $\epsilon$ is proportional to imaginary part of one of the complex parameters at play ($\mu_R$); ii) $\mu_{LR}$ as the other potentially complex parameter identified in Sec.~\ref{sec:lagrangian}, c.f.~\cref{eq:IIIA_scalarPotential}, plays no role here because its would-be phase is unphysical --- indeed, unlike that of $\mu_R$, it can be removed by a redefinition of the $\Phi_L$ field, see~\cref{eq:A1_scalarPotential}. 

Note that $\Delta_1$ also has $CP$-violating decays into fermions (c.f.~Fig.~\ref{fig:CPDeltafermi}); the $CPT$ theorem then ensures that the corresponding asymmetry is complementary to that in the $\Delta_1\to\Upsilon H_1^*$ and $\Delta_1^*\to\Upsilon^* H_1$ channels, such that the total decay widths of $\Delta_1$ and $\Delta_1^*$ remain equal. 

\subsection{Phase 2: Long-lived $\Upsilon$+RHN-driven washout}
\begin{figure}[t]
    \centering
    \raisebox{-0.5\height}{
    \includegraphics[scale=1.25]{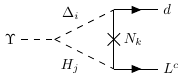}}\hspace{-1mm}
    \raisebox{-0.5\height}{
    \includegraphics[scale=1.25]{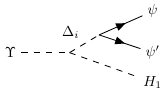}}
    \caption{Two- and three-body decays of $\Upsilon$ into SM fields (left and right, respectively).}
    \label{fig:Upsilondecay}
\end{figure}

To realise the mechanism described in Sec.~\ref{sec:mechanism}, one of the decay products of $\Delta_1$ needs to be long-lived and have only feeble interactions with the SM fermions, such that it can store the asymmetry temporarily. This role is played by $\Upsilon$, which, due to the Pati-Salam symmetry, does not have tree-level Yukawa couplings to the SM fermions. Instead, its couplings to $d$ and $L^c$ are realised at 1-loop-level, as seen in Fig.~\ref{fig:Upsilondecay} (left), and the corresponding width goes roughly as
\begin{equation}
        \Gamma\lr{\Upsilon\to d L^c} \propto  \frac{1}{(8\pi)^5}
        \frac
        {|\mu_{LR}|^2 m_\Upsilon^2 m_{N}}{m_{\Delta}^4}
        \log
        \lr{
        \frac{m_{\Delta}^2}{m_{N}^2}
        } \,,
\end{equation}
c.f.~\cref{eq:UpsilontoSMGamma}, where $m_N$ stands for the right-handed neutrino Majorana mass scale. 

Hence, for small enough $m_\Upsilon/m_\Delta$ and/or $m_N/m_\Delta$ and/or $\mu_{LR}/m_\Delta$  this channel is {\em very strongly} suppressed, and typically goes much slower than the three-body tree-level $\Upsilon$ decay shown in Fig.~\ref{fig:Upsilondecay} (right).
\begin{figure}[t]
    \centering
    \raisebox{-0.5\height}{
    \includegraphics[scale=1.25]{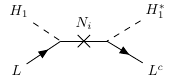}}
    \caption{Washout process mediated by $N$. In addition there are two $t$-channel diagrams, c.f.~Appendix~\ref{app:boltzmann}.}
    \label{fig:washout}
\end{figure}

Finally, the $\Delta L=2$ two-to-two scatterings mediated by RHNs, e.g.\ via the diagram shown in Fig.~\ref{fig:washout}, lead to a washout channel that acts on SM fermions while not affecting the asymmetry stored in $\Upsilon$--$\Upsilon^*$. Note that at $T\sim m_N$ there could be an additional source of asymmetry generation via $CP$-violating decays of $N$ (i.e. the usual thermal leptogenesis). We do not consider such effects here and take $N$-decays to be $CP$-conserving. 

\subsection{Boltzmann equations}
To obtain the set of Boltzmann equations corresponding to the present scenario we use the formalism described in Appendix~\ref{app:boltzmann}. For convenience we define a condensed set of equilibrium reaction rates
\begin{align}
    &\begin{aligned}\label{eq:gamma1}
    \gamma_D &\equiv  \gamma^\eq\!\lr{\Delta_1\to\Upsilon H_1^*}+\gamma^\eq\!\lr{\Delta_1\to \psi\psi'}\,,
    \end{aligned}\\
    &\begin{aligned}
    \!\gamma_{gg} &\equiv  \gamma^\eq\!\lr{\Delta_1\Delta_1^*\to gg}\,,
    \end{aligned}\\
    &\begin{aligned}
    \!\!\gamma_{\Upsilon}^{\text{tot}} &\equiv \gamma^\eq\!\lr{\Upsilon\to H_1\psi\psi'}+\gamma^\eq_\text{sub}\lr{\Upsilon H_1^*\to \psi\psi'}\\
    &+\gamma^\eq\!\lr{\Upsilon \psi^c\to H_1\psi'}\,,
    \end{aligned}\\
    &\begin{aligned}\label{eq:gamma4}
    \!\!\!\gamma_{N_W} &\equiv \gamma^\eq\!\lr{LH_1 \to \bar{L}H_1^*}+\gamma^\eq\!\lr{LL \to H_1^*H_1^*} .
\end{aligned}
\end{align}
In what follows we choose to express the relevant set of Boltzmann equations in terms of three important number densities (normalized to that of photons, $\eta_i\equiv n_i/n_\gamma$), namely
\begin{equation}
\begin{aligned}
    2\eta_{\Delta_1^+} &\equiv \eta_{\Delta_1}+\eta_{\Delta_1^*}\,,\\
    2\eta_{\Upsilon^-} &\equiv  \eta_{\Upsilon}-\eta_{\Upsilon^*}\,,\\
\end{aligned}
\end{equation}
and also $\eta_{B-L}^{\rm SM}$ denoting the density of the $B-L$ charge
stored in the SM fermions (i.e.\ all fermions besides RHNs). Note that these definitions ($\eta_{\Upsilon^-}$ deriving from particle {\em number} densities while $\eta_{B-L}^{\rm SM}$ corresponding to the $B-L$ {\em charge} density) mean that the exact balance between the $B-L$ asymmetry stored in the $\Upsilon-\Upsilon^*$ population and that stored in the SM (+RHNs) after $\Delta_1$ and $\Delta_1^*$ disappearance corresponds to $\eta_{\Upsilon^-}=\tfrac{3}{4} \eta_{B-L}^{\rm SM}$; see also~\cref{eq:BE3rd} and Appendix~\ref{app:boltzmann}.
With this at hand, one arrives at 
\begin{widetext}
\begin{align}
    zHn_\gamma \frac{d\eta_{\Delta_1^+}}{dz} = &-\gamma_D\Bigg[\left(\frac{\eta_{\Delta_1^+}}{\eta_{\Delta_1}^\eq}-1\right)
    +\epsilon\lr{\frac{\eta_{\Upsilon^-}}{\eta_{\Upsilon}^\eq}+\eta_{B-L}^{\rm SM}}\Bigg]-\gamma_{gg}\Bigg[\lr{\frac{\eta_{\Delta_1^+}}{\eta_{\Delta_1}^\eq}}^2-1\Bigg]\label{eq:BE1st}]\,,\\
    zHn_\gamma \frac{d\eta_{\Upsilon^-}}{dz} = & -\gamma_D\epsilon\left(\frac{\eta_{\Delta_1^+}}{\eta_{\Delta_1}^\eq}-1\right)-\Big(\gamma_{\Upsilon}^\text{tot}+\gamma_Dr^2\Big)\frac{\eta_{\Upsilon^-}}{\eta_{\Upsilon}^\eq}-\Big(\gamma_{\Upsilon}^\text{tot}-\gamma_Dr(1-r)\Big)\eta_{B-L}^{\rm SM}\label{eq:BE2nd}\,,\\
    \frac{3}{4}zHn_{\gamma}\frac{d\eta_{B-L}^{\rm SM}}{dz} = & -\gamma_D\epsilon\left(\frac{\eta_{\Delta_1^+}}{\eta_{\Delta_1}^\eq}-1\right)-\Big(\gamma_{\Upsilon}^\text{tot}-\gamma_Dr(1-r)\Big)\frac{\eta_{\Upsilon^-}}{\eta_{\Upsilon}^\eq}-\Big(\gamma_{\Upsilon}^\text{tot}+\gamma_D(1-r)^2+\gamma_{N_W}\Big)\eta_{B-L}^{\rm SM}\, . \label{eq:BE3rd}
\end{align}
\end{widetext}
with the interaction rate factors ($\gamma$'s) defined in Eqs.~(\ref{eq:gamma1}) to~(\ref{eq:gamma4}). 
Note that~\cref{eq:BE1st,eq:BE2nd,eq:BE3rd} represents a condensed form of the full set of evolution equations, c.f.~\cref{eq:Dpeq,eq:BLeq}, with the $\eta_N$ and $\eta_{\Delta_1^-}$ sectors omitted for simplicity. As discussed in Appendix~\ref{app:boltzmann}, this modification should have only minor impact on the final yields; nevertheless, the numerical analysis of Sec.~\ref{sec:results} has been performed without this simplification.

\begin{figure*}
    \centering
    \includegraphics[width=0.65\linewidth]{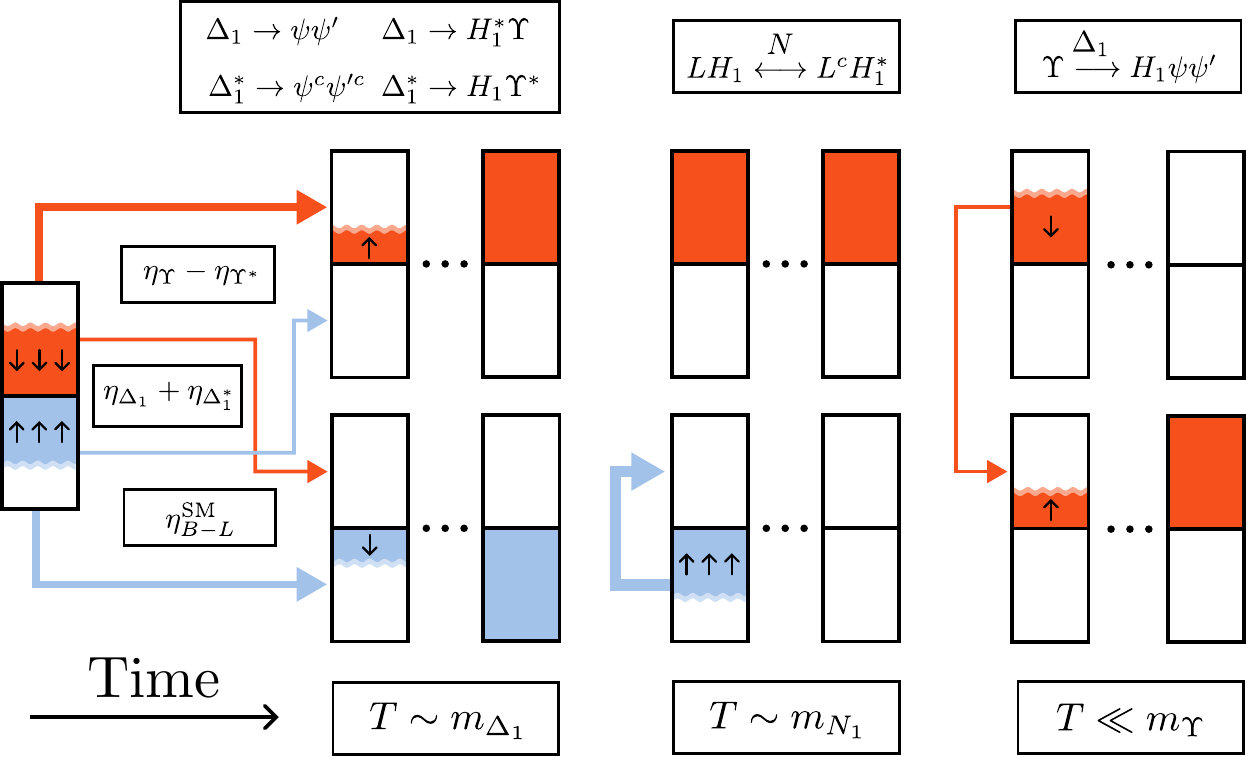}
    \caption{
    Illustration depicting the baryogenesis mechanism described in the text, where a final baryon asymmetry can arise despite a period of strong washout after its initial generation. 
    At $T\sim m_{\Delta_1}$, $CP$-violating decays of $\Delta_1$ and $\Delta_1^*$ lead to asymmetries in both the $\Upsilon$--$\Upsilon^*$ ($\eta_{\Upsilon^-}$) and the SM matter ($\eta_{B-L}^{\rm SM}$) sectors . Later, assuming $m_{\Delta_1}>m_N$, the washouts mediated by $N$ are strongest around $T\sim m_N$, at which time they may completely or partially erase $\eta_{B-L}^{\rm SM}$. Even later, at some $T\ll m_\Upsilon$ when $\Gamma_{\Upsilon}\sim H$ (assuming a small decay width $\Gamma_\Upsilon$), the asymmetry stored in the $\Upsilon$--$\Upsilon^*$ sector gets transferred into the SM.
    }
    \label{fig:BLsmugglingdrawing}
\end{figure*}

In~\cref{eq:BE1st,eq:BE2nd,eq:BE3rd}, the baryogenesis mechanism we consider is then realized as follows:
\begin{enumerate}
    \item When $\Delta_1$ goes out of equilibrium, the decay rate $\gamma_D$ generates the $\eta_{\Upsilon^-}$ and $\eta_{B-L}^{\rm SM}$ densities, both proportional to $-\epsilon$ via the first terms in \cref{eq:BE1st,eq:BE2nd,eq:BE3rd} (i.e.~the first round brackets). 
    Furthermore, both Eqs.~\eqref{eq:BE2nd} and~\eqref{eq:BE3rd} contain identical washout terms proportional to $\gamma_{\Upsilon}^\text{tot}$. 
    With only these terms, one would expect the evolution of $\eta_{\Upsilon^-}$ and $\eta_{B-L}^{\rm SM}$ to be the same (up to the factor $3/4$), such that, when $\Upsilon$ eventually decays, the net $B-L$ in SM fermions zeroes out again unless this balance is disturbed in some way.
    
    Note that this effect can also be inferred from the diagram in Fig.~\ref{fig:Upsilondecay} (right), being the dominant decay mode of $\Upsilon$, where we see that the eventual $\Upsilon$ decays lead to the same overall final state as $\Delta_1$ decays to fermions (compare $\Delta_1\to \psi\psi'$ with $\Delta_1\to \Upsilon H^*$ followed by $\Upsilon\to H\psi\psi'$). Hence, without further processing on either side, any $B-L$-asymmetry generated in $\Delta_1$-decays would eventually get cancelled by the $B-L$-asymmetry generated in the decays of $\Upsilon$, which is equal in magnitude but opposite in sign.
    \item 
    However, if some process changes either $\eta_{\Upsilon^-}$ or $\eta_{B-L}^{\rm SM}$ before $\Upsilon$ decays, this cancellation will not be exact. In particular, the term proportional to $\gamma_{N_W}$ in Eq.~\eqref{eq:BE3rd} encodes the effects of the $N$-mediated washout processes that, depending on the mass and interaction strength of RHNs, can completely or partially wash out the $B-L$-asymmetry in SM leptons (and, in collaboration with sphalerons, in all SM fermionic species), without affecting $\eta_{\Upsilon^-}$ of the long-lived $\Upsilon$ and $\Upsilon^*$ populations. 
    \item
    When the long-lived $\Upsilon$'s eventually decay, the asymmetry stored therein then gets transferred to the SM fermions without cancellation, and the $B-L$-asymmetry is restored.  
\end{enumerate}
A simple pictorial representation of these three basic stages is given in Fig.~\ref{fig:BLsmugglingdrawing}.

Interestingly, even in the (theoretical) absence of $N$-mediated washout, a difference between $\eta_{\Upsilon^-}$ and $\eta_{B-L}^{\rm SM}$ can still emerge, owing to the $\gamma_D$-proportional terms in~\cref{eq:BE2nd,eq:BE3rd} (differing by $r$ vs. $(1-r)$ factors, respectively). For $r\neq \frac{1}{2}$ this leads to an unequal washout between $\eta_{\Upsilon^-}$ and $\eta_{B-L}^{SM}$ via inverse decays into $\Delta_1$, again breaking the degeneracy between the two asymmetries.
Note also that, technically, for very small ratios of $\Gamma_{\Upsilon}/m_\Upsilon$ the Universe may enter a period of matter domination before $\Upsilon$ decays. We ignore the nuances introduced by such a twist and simply assume radiation domination throughout all parameter space scans discussed in the next section.

%
\section{Sample numerical analysis\label{sec:numerics}}
%
\subsection{Parameters and benchmark points\label{sec:parametersbenchmarks}}
The model described in Sec.~\ref{sec:patisalam} contains a large number of free parameters such as mass scales and coupling constants. For simplicity, we choose to present the results in terms of a reduced set of parameters, which we obtain by imposing specific relations on the different mass scales. Note however that the breadth of the viable parameter space is not limited to these considerations from a model perspective. 

Firstly, we associate several mass scales to the Pati-Salam breaking scale $\langle \Phi_R^{42}\rangle\equiv V_G$, such that
\begin{equation}
\label{eq:constraint1}
    V_G = m_\Omega = m_\Sigma\, .
\end{equation}
Next, we introduce a second lower scale $\overline{\mu}$ such that
\begin{equation}
\label{eq:constraint2}
    \overline{\mu} = \re\lr{\mu_R} = \im\lr{\mu_R} = \mu_{LR} = m_L\, .
\end{equation}
This choice simultaneously renders a light scalar $H_1$ (with $\overline{\mu}\gg m_{H_1}$) and opens the decay channel $\Delta_1\to \Upsilon H_1$, constituting one of the ways in which the baryogenesis mechanism described in Sec.~\ref{sec:mechanism} can be realised. Note that having a light $H_1$ in this context goes against the extended survival hypothesis~\cite{Babu:2015bna} and could be argued to require fine-tuning. However, the baryogenesis mechanism itself only requires $m_\Upsilon \gtrsim m_{H_1}$ and does not rely on this tuning. Note further that having $\re\lr{\mu_R} = \pm\im\lr{\mu_R}$ maximises the amount of $CP$-violation in $\Delta_1$-decays. 

Next, we consider the Yukawa couplings to be scalar matrices in flavour space such that
\begin{equation}
\label{eq:constraint3}
    Y_{R (LR)} = \textsl{y}_{R (LR)}\mathbb{1}_{3\times 3}\, ,
\end{equation}
where $\textsl{y}_{R (LR)}$ is a scalar. Note that this structure is used out of simplicity, and that matching $\textsl{y}_{LR}$ to the SM Yukawa couplings requires e.g.\ an additional Pati-salam multiplet containing a Higgs-like scalar, or careful consideration of the effective operators described in Appendix~\ref{app:fullmodel}. 

From Eq.~\eqref{eq:constraint3} we find a left-handed neutrino mass scale $m_\nu=\textsl{y}_{LR}^2v^2m_N^{-1}$, where $m_N$ is the right-handed neutrino mass, which we use to match with Eq.~\eqref{eq:mnurange}. Finally, the quartic couplings $\lambda_5$ and $\lambda_6$ are already constrained by Eq.~\eqref{eq:constraint4}, and the gauge couplings $g_4$ and $g_R$ are taken as the values that match with the SM hypercharge gauge coupling evolved to $V_G$, c.f.\ Eq.~\eqref{eq:gYgc}. 

We stress that \cref{eq:constraint1,eq:constraint2,eq:constraint3} represent choices made in order to effectively study the parameter space, not constraints inherent in the model or essential for the baryogenesis mechanism to generate a sizeable asymmetry. 

Given this, we specify a couple of benchmark points BM1 and BM2 as
\begin{center}
\begin{tabular}{|clll|}
     \hline 
     \multirow{2}*{\textbf{BM1}: } & $V_G=10^{15}$~GeV, & $\textsl{y}_R=10^{-1}$, & $\lambda_5=-10^{-1}$, \\
     & $\overline{\mu}=2\times 10^{-2}\, V_G$, & $\textsl{y}_{LR}=10^{-1}$, & $\lambda_6=1$\, ; \\
     \hline
\end{tabular}
\end{center}
and
\begin{center}
\begin{tabular}{|clll|}
     \hline 
     \multirow{2}*{\textbf{BM2}: } & \multirow{2}*{BM1 \;+\; } & $ C_{LR2}={Y_{LR}}/{V_G}\,$, & \\
     &  & $C_{RR2}=10^{-2}\times Y_R/V_G\,$ , & \\
     \hline
\end{tabular}
\end{center}
on which we shall illustrate the salient features of the setting. 

Note that the latter differs from the former by including the pair of dimension-5 effective operators $\mathcal{O}_{LR2}$ and $\mathcal{O}_{RR2}$ of Eq.~\eqref{eq:ops12}. 
The main phenomenological difference between the two scenarios then stems from the fact that the presence of $\mathcal{O}_{LR2}$ within BM2 enhances the coupling of the $H_1$ mass eigenstate to the $N\bar{L}$ fermionic bilinear by providing (tree-level forbidden, absent in BM1) direct coupling to its dominant $H'$ component. 
At the same time, $\mathcal{O}_{RR2}$ gives an additional bare Majorana mass contribution $m_N^0~\sim 10^{12}$~GeV to $N$ (to be compared with $m_N\sim 4\times 10^7$~GeV for BM1 via the two-loop diagram in Fig.~\ref{fig:twoloopMN}). 

With these Wilson coefficients, the contributions of $\mathcal{O}_{LR2}$ and $\mathcal{O}_{RR2}$ 
conspire to approximately cancel each other in the type-I seesaw mechanism, leaving the light neutrino mass $m_\nu$ largely unchanged. However, the increase in the strength of the $N$--SM interaction leads to a stronger washout, and the increased mass of $N$ means that this washout has an earlier onset. Note further that, for this benchmark point, the additional coupling between $\Delta_1$ and the $\bar{d^c}N$ fermion bilinear generated via $\mathcal{O}_{RR2}$ is of the same order as the one stemming from the tree-level Yukawa coupling (c.f.\ Sec.~\ref{sec:lagrangian}). 

Note that for some values of the parameters discussed above the mass of $\Delta_1$ can become comparable to the (heavy) gauge boson masses. However, we make the simplifying assumption that the gauge degrees of freedom do not play any significant role in the baryogenesis mechanism of our interest, which can easily be achieved by simultaneously scaling down all scalar mass parameters with respect to $V_G$. 
Furthermore, we take the initial number density of $\Delta_1$ to be in equilibrium while assuming all other initial number densities  to be zero. 

\begin{figure*}
    \centering
    \includegraphics[width=0.49\linewidth]{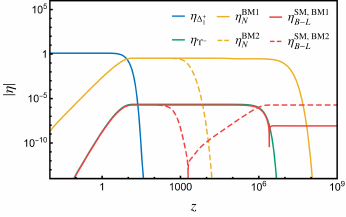}
    \includegraphics[width=0.49\linewidth]{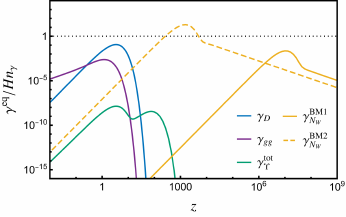}
    \caption{\textbf{Left:}
    Number densities $\eta_i\equiv n_i/n_\gamma$ of different particle species $i$ (normalised to the photon number density $n_\gamma$) as functions of $z\equiv m_{\Delta_1}/T$, for benchmark points BM1 (solid lines) and BM2 (dashed lines) described in Sect.~\ref{sec:parametersbenchmarks}. \textbf{Right:} Equilibrium reaction rates $\gamma^\eq$, normalised to $Hn_\gamma$, where $H$ is the Hubble rate, as functions of $z$. 
    %
    Note that the fact that $T\sim M_N$ occurs only after $\Upsilon$ and $\Upsilon^*$ decay ($\eta_N$ remains large at their onset) in the BM1 case does not mean that the asymmetry transferred by $\Upsilon$'s into the SM gets simply washed out; the key is that the RHN-driven washout rate at this point remains small compared to the Hubble rate, see the right figure. In contrast to that, in the BM2 setting the RHN-driven washout becomes efficient for a while (the bump on the dashed-yellow curve in the right figure makes it over 1 for a short while), but this dates to the era when $\Upsilon$ still did not start to decay; hence, the asymmetry stored therein remains safe.}
    \label{fig:lineplots}
\end{figure*}

\subsection{Results} \label{sec:results}

\begin{figure*}[ht!]
    \centering
    \includegraphics[width=0.44\linewidth]{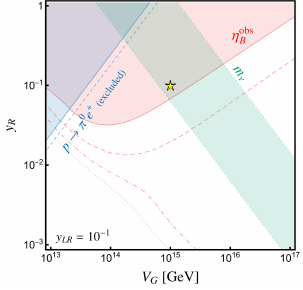}\hspace{0.05\linewidth}
    \includegraphics[width=0.44\linewidth]{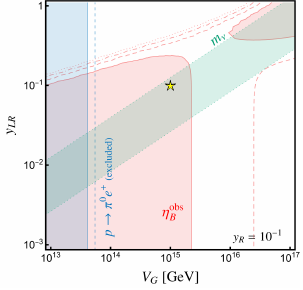}
    \caption{\textbf{Left:} Baryon asymmetry normalized to photon density ($\eta_{B}$, in red) as a function of the Pati-Salam symmetry-breaking scale $V_G$ as well as the Yukawa coupling $\textsl{y}_R$, for the parameter choices in \cref{eq:constraint1,eq:constraint2,eq:constraint3} along with the specific values $\overline{\mu}=2\times 10^{-2}\, V_G$ and $\textsl{y}_{LR}=10^{-1}$, as well as $\lambda_5=-10^{-1}$ and $\lambda_6=1$. The red area shows where a baryon asymmetry equal to or greater than the observed value $\eta_{B}^\text{obs}= 6.20\times 10^{-10}$ is generated, while the dashed, dot-dashed and dotted lines show $\eta_{B}$ a factor of $10^{-1}$, $10^{-2}$ and $10^{-3}$ below $\eta_{B}^\text{obs}$, respectively. The blue area is excluded by searches for the proton decay mode $p\to \pi^0e^+$ at Super-Kamiokande, with the dashed line showing the projected future reach at Hyper-Kamiokande. The green area shows where the neutrino mass scale  generated via type-I seesaw, c.f.\ Sec.~\ref{sec:phenomnu} falls into the observed ballpark. The yellow star shows the location of benchmark point BM1, as described in the text. \textbf{Right:} Same as \textbf{left}, but as a function of $\textsl{y}_{LR}$ with $\textsl{y}_{R}=10^{-1}$.}
    \label{fig:VGcontour}
\end{figure*}

\begin{figure*}
    \centering
    \includegraphics[width=0.44\linewidth]{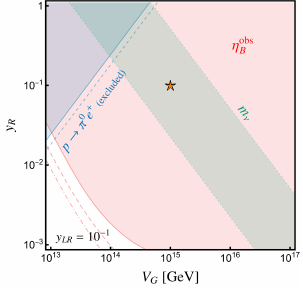}\hspace{0.05\linewidth}
    \includegraphics[width=0.44\linewidth]{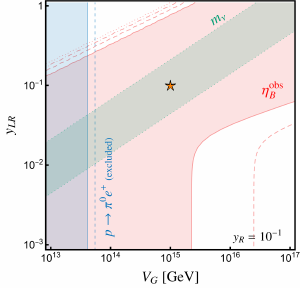}
    \caption{\textbf{Left:} Same as Fig.~\ref{fig:VGcontour} (\textbf{left}), but with the inclusion of two dimension-5 effective operators $\mathcal{O}_{LR2}$ and $\mathcal{O}_{RR2}$ with Wilson coefficients according to benchmark point BM2 (see Sect.~\ref{sec:parametersbenchmarks} for details). The orange star marks BM2. \textbf{Right:} Same as \textbf{left}, but as a function of $\textsl{y}_{LR}$ with $\textsl{y}_{R}=10^{-1}$.}
    \label{fig:VGcontourOp}
\end{figure*}

In Fig.~\ref{fig:lineplots} (left), we show, for benchmark points BM1 (solid lines) and BM2 (dashed lines), the evolution of the absolute value of the number densities $\eta_i\equiv n_i/n_\gamma$ of different particle species or combinations thereof, with respect to $z\equiv m_{\Delta_1}/T$, where $T$ is the temperature. Note that for $\eta_{\Delta_1^+}$ and $\eta_{\Upsilon^-}$, the evolution of the densities is identical between the two benchmark points. We see that, as $\Delta_1$ decays (blue line), 
the $\eta_{\Upsilon^-}$, $\eta_{N}$ and $\eta_{B-L}^{\rm SM}$ densities are populated (yellow, green and red lines, respectively), and subsequently freeze out once $\eta_{\Delta^+}$ has significantly dropped.

For BM1 (solid lines), when $\eta_{\Upsilon^-}$ later drops, $\eta_{B-L}^{\rm SM}$ flips sign and freezes out at a lower value. This feature comes from a near-cancellation between contributions to $\eta_{B-L}^{\rm SM}$ generated in $\Delta_1$ and $\Upsilon$ decays, respectively, having opposite signs and being near-equal in magnitude. They are close to equal since, as discussed in Sec.~\ref{sec:baryogenesis}, $\Delta_1$-decays generate one unit of $\eta_{\Upsilon^-}$ for every $4/3$ units of $\eta_{B-L}^{\rm SM}$, and each subsequent $\Upsilon$-decay generates $-4/3$ units of $\eta_{B-L}^{\rm SM}$, effectively cancelling the asymmetry from $\Delta_1$. However, for benchmark point BM1, washout via inverse decay into $\Delta_1$ is much more rapid via the fermion channel ($\psi \psi'\to\Delta_1$) compared to the scalar channel ($\Upsilon H_1^* \to \Delta_1$), corresponding to $r\to 0$ (c.f.\ Sec.~\ref{sec:baryogenesis}). Washout via inverse decay is independent of the $CP$-violating parameter $\epsilon$, and leads to an overall reduction of $\eta_{B-L}^{\rm SM}$ while not significantly affecting $\eta_{\Upsilon^-}$. 
Therefore, during $\Delta_1$-decays, $\eta_{B-L}^{\rm SM}$ is slightly washed out while $\eta_{\Upsilon^-}$ remains untouched, breaking the degeneracy between them and leading (much later) to a non-exact cancellation of $\eta_{B-L}^{\rm SM}$ when $\Upsilon$ decays. We see also in BM1 that $\eta_{N}$ drops later than $\eta_{\Upsilon^-}$, such that the $N$-mediated washouts are strongest after $\eta_{B-L}^{\rm SM}$ has frozen out for the second time. For large couplings between $N$ and the SM this could have removed the $\eta_{B-L}^{\rm SM}$ asymmetry completely, but in this benchmark point the $N$-mediated washouts are sufficiently weak for the asymmetry to survive. 

The dashed green and red lines in Fig.~\ref{fig:lineplots} (left) show the evolution of $\eta_{N}$ and $\eta_{B-L}^{\rm SM}$, respectively, in benchmark point BM2, where two dimension-5 effective operators 
are added to the Lagrangian, c.f.\ Sec.~\ref{sec:lagrangian}.
While the choice of Wilson coefficients in BM2 leaves the light neutrino mass almost unchanged, the effect on baryogenesis is significant, as can be seen by comparing the solid and dashed green and red lines in Fig.~\ref{fig:lineplots} (left). 
We see that the increased strength and earlier onset of the $N$-mediated washouts results in the $\eta_{B-L}^{\rm SM}$ asymmetry being washed out around $z\sim10^3$. However, at later times when $\Upsilon$ decays, the $\eta_{\Upsilon^-}$ asymmetry gets transferred to the SM without significant cancellation. In this way, $\Upsilon$ has protected the asymmetry generated in $\Delta_1$-decays, ``smuggled'' it past a period of strong washout mediated by $N$, and released the asymmetry to the SM once the washout has become weak.

We find the freeze-out value of the baryon asymmetry to be $\eta_B\approx 1.2\times 10^{-9}$ for BM1 and $\eta_B\approx 2.3\times 10^{-8}$ in BM2. Using the approximation in Eq.~\eqref{eq:etaBapproximation}, we find $\eta_B\approx 2.8\times 10^{-8}$ for BM2. Note that for this approximation we assumed that $N$-mediated washouts occur before $\Upsilon$-decays, which makes it incompatible with BM1.

For the same benchmark points BM1 and BM2 we show, in Fig.~\ref{fig:lineplots} (right), the corresponding reaction rate densities $\gamma^\eq$, normalized to $Hn_\gamma$, where $H$ is the Hubble rate. Note that $\gamma_\Upsilon^\text{tot}$ peaks twice: at $T\sim m_{\Delta_1}$ due to resonant scatterings, and again at $T\sim m_\Upsilon$ when the decay process is enhanced. However, the actual $\Upsilon$-decays don't occur until much later, since the couplings to SM fields is suppressed. Note further that only $\gamma_N$ is affected by the addition of the operators (solid vs.\ dashed yellow lines).

In Fig.~\ref{fig:VGcontour} (left) we show the final baryon asymmetry $\eta_{B}$ in the plane spanned by $\textsl{y}_R$ and $V_G$, with the rest of the parameters given by BM1, where $\eta_{B}\approx \frac{1}{79}\eta_{B-L}^{\rm SM}$ at $z\gg 1$. We see that successful baryogenesis (red areas) is possible without being ruled out by proton decay searches (blue). Furthermore, there is a significant overlap between the successful regions and the part of parameter space where the observed neutrino mass scale is generated via type-I seesaw (shown in green). However, note that this latter region may shift to higher or lower values of $V_G$ depending on the details of the flavour matching in the model, which we have not attempted to perform in this work, but which could potentially be achieved using the effective operators described in Appendix~\ref{app:fullmodel}. The region below the green area is excluded as it generally corresponds to overly large neutrino masses, but it may as before open up with flavour matching. 

For small values of $V_G$ and $\textsl{y}_R$ we see that the final baryon asymmetry rapidly drops (bottom-left region of Fig.~\ref{fig:VGcontour} left). This is due to the RH neutrinos $N$ becoming lighter, with $m_N\propto \textsl{y}_RV_G$, leading to $N$-mediated washouts being active at later times, when they are stronger with respect to $Hn_\gamma$ (note that for $T\gtrsim m_N$ we have $(\gamma_N/Hn_\gamma)\propto T^{-3}$, while the maximum value of $\gamma_N$ is found at $T\sim m_N$, such that $\gamma_N$ is effectively stronger for small $m_N$). For large $m_N$ (upper-right part of Fig.~\ref{fig:VGcontour} left) the $N$-mediated washouts are not strong enough to have any effect on $\eta_{B}$. Instead, this region of parameter space is dominated by the amount of washout $\eta_{B-L}^{\rm SM}$ experiences during the period of $\Delta_1$-decays, where higher values of $\textsl{y}_R$ and lower values of $V_G$ increase the strength of the inverse decay $\psi \psi'\to\Delta_1$, thereby breaking the degeneracy between $\eta_{B-L}^{\rm SM}$ and $\eta_{\Upsilon^-}$. For smaller $\textsl{y}_R$ and larger $V_G$, the inverse decays are weaker, leading to a greater degree of cancellation between the two.

In Fig.~\ref{fig:VGcontour} (right) we show the final baryon asymmetry as a function of $\textsl{y}_{LR}$ and $V_G$. Here we see that, for low enough values of $\textsl{y}_{LR}$, the final baryon asymmetry $\Delta B$ is independent of $\textsl{y}_{LR}$, as the $N$-mediated washouts are too weak to have any effect. Instead, the lower regions are completely dominated by the variation in $V_G$, where higher values again lead to a lower final asymmetry due to the effective reduction in the rate of inverse decay of fermions into $\Delta_1$, such that the degeneracy between $\eta_{B-L}^{\rm SM}$ and $\eta_{\Upsilon^-}$ is maintained to a larger extent. For high values of $\textsl{y}_{LR}$, the $N$-mediated washout is strong, leading to a complete removal of the asymmetry in case $\Upsilon$ decays before $N$ (top-left). The successful region in the top-right corner is due to $\Upsilon$ decaying after the strong $N$-mediated washouts have already erased the temporary $\eta_{B-L}^{\rm SM}$, such that there is no cancellation when the $\eta_{\Upsilon^-}$ asymmetry is eventually transferred to the SM. Note that while low values of $V_G$ are excluded by proton decay searches, such constraints could be to a large degree avoided in the absence of first-generation couplings in the BSM sector, and the scale of baryogenesis could in principle be quite low.

In Fig.~\ref{fig:VGcontourOp} (left) we show the same parameter space scan as in Fig.~\ref{fig:VGcontour} (left), but with the addition of two dimension-5 effective operators, $\mathcal{O}_{LR2}$ and $\mathcal{O}_{RR2}$, with Wilson coefficients according to benchmark point BM2. Here we see that successful baryogenesis is realised for a large part of the parameter space. The effective operators lead to an increased strength of the $N$-mediated washout, as well as an earlier onset, such that $\eta_{B-L}^{\rm SM}$ is washed out before $\Upsilon$ decays. This allows $\eta_{\Upsilon^-}$ to hide the asymmetry and repopulate it at later times without cancellation. For small values of $\textsl{y}_R$ and $V_G$, the decreased RH neutrino mass $m_N$ leads to a later onset of $N$-mediated washouts, potentially pushing it beyond the time of $\Upsilon$-decays and thereby disabling the degeneracy breaking (bottom left).

In Fig.~\ref{fig:VGcontourOp} (right) we similarly show how adding $\mathcal{O}_{LR2}$ and $\mathcal{O}_{RR2}$ according to BM2 affects the viable parameter space when varying $\textsl{y}_{LR}$. For large $\textsl{y}_{LR}$ and small $V_G$, the $N$-mediated washout is still strong after $\Upsilon$ decays, such that no final asymmetry survives. However, for small $\textsl{y}_{LR}$, the $N$-mediated washouts are not strong enough to affect $\eta_{B-L}^{\rm SM}$, and the magnitude of the final asymmetry is instead completely determined by $V_G$, where smaller values again lead to increased inverse decay of fermions into $\Delta_1$, thereby breaking the degeneracy between $\eta_{B-L}^{\rm SM}$ and $\eta_{\Upsilon^-}$. Between these two regimes the $N$-mediated washouts are strong, as well as being active before $\Upsilon$-decays, leading to an increased final baryon asymmetry by washing out the $\eta_{B-L}^{\rm SM}$ at intermediate temperatures.

\section{Conclusion}\label{sec:conclusion}
We present a simple unified model based on the Pati-Salam $SU(4)_C\times SU(2)_L\times SU(2)_R$ symmetry, in which a ``primordial'' $B-L$-asymmetry generated by high-scale decays of colour triplet scalars can be protected from subsequent strong $\Delta L=2$ washout inherent to leptogenesis scenarios, if it is temporarily stored in a relatively long-lived BSM sector essentially decoupled from the RH neutrinos, and only later released to the SM. 

Unlike in previous works, this scenario does not rely on avoiding the specific time-window of rapid electroweak sphaleron transitions, or on the initial generation of the asymmetry in a specific lepton flavour. Instead, we find that the interplay of $B-L$-violating scalar decays and RHN-driven processes naturally leaves large regions of the relevant parameter space viable for baryogenesis, owing to the non-trivial pattern of scalar couplings to the SM enforced by the Pati-Salam symmetry. This may also be viewed as a counter-example to the usual expectation that the presence of efficient low-scale $B-L$ violation rather generally rules out high-scale baryogenesis.

The details of this attractive dynamical picture are illustrated in two benchmark scenarios differing by the set of higher-dimensional operators invoked, without any specific assumptions imposed on the relevant Yukawa couplings. We check that the type-I seesaw, driven by RH neutrino masses generated at two loops, provides a reasonable light neutrino spectrum in both cases, and that all constraints from proton instability inherent to the current scenario are satisfied.    

While the model was kept minimal for simplicity, we note that several additional interesting effects may arise from the couplings that were deliberately omitted. In particular, the presence of two separate tri-linear couplings that both involve the relevant scalar fields may generally allow the asymmetry to be fully generated in the scalar sector, even in the absence of tree-level Yukawa couplings. In our present Pati-Salam scenario, this may be realized by the inclusion of the tri-linear coupling $\mu_L$ (c.f.\ Appendix~\ref{app:fullmodel}), allowing $\Delta_1$ to decay into a pair of $\Upsilon^*$'s. This then provides a second decay mode as required by $CPT$-conservation, without relying on the coupling between $\Delta_1$ and the SM fermions.

Another example of a model in which the mechanism studied in this work may be naturally implemented is the minimal flipped $SU(5)\times U(1)$ scenario with radiatively generated seesaw scale~\cite{ArbelaezRodriguez:2013kxw,Harries:2018tld,Fonseca:2023per,Malinsky:2025jfu}, to be scrutinized in a future work. 


\begin{acknowledgements}
The authors acknowledge the use of FeynCalc~\cite{Shtabovenko:2023idz} and FeynArts~\cite{Hahn:2000kx}. K.F.\ and M.M.\ acknowledge support from the FORTE project CZ.02.01.01/00/22 008/0004632 co-funded by
the EU and the Ministry of Education, Youth and Sports of the Czech Republic. K.F.\ additionally  acknowledges support from the Swedish Research Council (Dnr. 2022-04299). J.P-S. acknowledges support from the grants CIACIF/2022/158 and CIBEFP/2026/108, funded by Generalitat Valenciana (Spain). The work has also been supported by the Charles University Research Center of Excellence grant UNCE/24/SCI/016.
\end{acknowledgements}

\appendix
\section*{APPENDIX}
\section{Model details}\label{app:fullmodel}
In this Appendix, we discuss some of the technical details of the Pati-Salam model described in Sec.~\ref{sec:patisalam}.\\

\subsection{The Pati-Salam group and algebra \label{app:patisalam}}
The $SU(4)_C$ ``hypercolor'' part of the $SU(4)_C\times SU(2)_L\times SU(2)_R$ Pati-Salam group has a total of $4^2-1=15$ generators, with $4-1=3$ of them simultaneously diagonalizable. In the fundamental representation, these are all $4\times 4$ hermitean traceless matrices. The first eight generators can be defined by embedding the fundamental (triplet) generators of the $SU(3)_c$ color subgroup of $SU(4)_C$ within this set:
\begin{equation}\label{eq:T1to8}
\begin{aligned}\setlength\arraycolsep{1.5mm}
    T^a_4 = \frac{1}{2}\left(\begin{array}{ccc|c}
     &  &  & 0  \\
     & \scalebox{1.25}{$\lambda^a$} &  & 0  \\
     &  &  & 0 \\
    \hline
    0 & 0 & 0 & 0  \\
    \end{array}\right)
\end{aligned}
\end{equation}
where $\lambda^a$ with $a=1,2,\ldots,8$ are Gell-Mann matrices. Together, these matrices form an $\mathfrak{su}(3)$ subalgebra within the $\mathfrak{su}(4)$ algebra. Six of the seven remaining generators are the usual block-off-diagonal analogues of the off-diagonal Pauli matrices ($\sigma_1$ and $\sigma_2$) of $\mathfrak{su}(2)$,
\begin{equation}
\begin{aligned}\setlength\arraycolsep{1.5mm}
    T^a_4 = \frac{1}{2}\left(\begin{array}{ccc|c}
     &  &  &  \\
     & \scalebox{1.25}{$0_{3\times 3}$} &  & \vec{v}^\dagger  \\
     &  &  &  \\
    \hline
     & \vec{v} & & 0  \\
    \end{array}\right)\, ,
\end{aligned}
\end{equation}
 where $\vec{v}\equiv\big(\delta_{a9}+i\delta_{a10}, \; \delta_{a11}+i\delta_{a12}, \; \delta_{a13}+i\delta_{a14}\big)$, for $a=9,10,\ldots,14$. The remaining diagonal generator 
 \begin{equation}\label{eq:T15}
T_{15} = \frac{1}{2\sqrt{6}}\left(\begin{array}{cccc}
    1 & 0 & 0 & 0 \\
    0 & 1 & 0 & 0 \\
    0 & 0 & 1 & 0 \\
    0 & 0 & 0 & -3
\end{array}\right)\,.
\end{equation}
proportional to the physical $B-L$ charge, is then fixed by the standard normalization condition \begin{equation}
    \operatorname{Tr}\Big\{T^aT^b\Big\} = \frac{1}{2}\delta^{ab}.
\end{equation}
We follow the common practice of writing $SU(N)$ representations as tensors with upper (lower) indices for the fundamental (antifundamental) representations. Index contractions can only involve an upper and a lower index, $A^{\alpha\beta i} B_{\beta i}\equiv C^\alpha$. Complex conjugation raises/lowers all indices of a representation, $A_\alpha^{\;\, i} = (A^\alpha_{\;\, i})^*$. 

The $SU(4)_C$ representations employed in the current study are the following:

\begin{itemize}
    \item The fundamental $\mathbf{4}$: We write it as a tensor with an upper index $A^\alpha$, $\alpha=1,2,3,4$.
    \item The antifundamental  $\mathbf{\overline{4}}$: Same as the fundamental but with a lower index, $A_\alpha$ with $\alpha=1,2,3,4$.
    \item The 2-index antisymmetric tensor $\mathbf{6}$:
    We write it as a matrix with two upper fundamental indices $A^{\alpha\beta}$, satisfying $A^{\alpha\beta}=-A^{\beta\alpha}$. It is a real representation  equivalent to its dual, $\epsilon^{\alpha\beta\gamma\delta} A_{\gamma\delta}\equiv A^{\alpha\beta}$, where $\epsilon^{\alpha\beta\gamma\delta}$ is the $SU(4)_C$-invariant totally antisymmetric symbol.
    \item The adjoint $\mathbf{15}$:
    It is written as a tensor with upper and lower fundamental indices $A^{\alpha}_{\;\,\beta}$ satisfying $A^{\alpha}_{\;\,\alpha}=0$\;. It is a real representation. 
\end{itemize}
The $SU(4)_C$ fundamental representation decomposes into irreducible $SU(3)_c$ components as:
\begin{equation}
    \mathbf{4}\to \mathbf{3}\oplus\mathbf{1}\;,
\end{equation}
where, in the convention of Eq.~(\ref{eq:T1to8}) the triplet spans the first three components and the singlet corresponds to the fourth. Similarly, the decompositions of $\mathbf{6}$ and $\mathbf{15}$ are readily identified:
\begin{equation}
    \mathbf{6} \to \mathbf{3}\oplus\mathbf{\overline{3}}\;,\quad \mathbf{15} \to\mathbf{8}\oplus\mathbf{3}\oplus\overline{\mathbf{3}}\oplus\mathbf{1}\;.
\end{equation}

Concerning the $SU(2)_L\times SU(2)_R$ part of the Pati-Salam group (with the first factor  conventionally identified with the SM weak isospin), we follow a similar upper and lower index convention as before and 
employ two different set of indices ($i,j,k,\ldots$ for $SU(2)_L$ and $a,b,c,\ldots$ for $SU(2)_R$) to distinguish between the two sectors.

A non-zero VEV of the SM singlet in $\Phi_R= (\mathbf{\overline{4}},\mathbf{1},\mathbf{2})$ of Pati-Salam corresponding to its $(\Phi_R)_4^{\;2}$ component breaks the $SU(4)_C\times SU(2)_R$ subgroup of Pati-Salam into $SU(3)_c\times U(1)_Y$ of the SM, with the latter factor generated by
\begin{equation}
    Y = T_R^3 + \sqrt{\frac{2}{3}}\,T_4^{15}\;.
\end{equation}

\subsection{Scalar potential\label{app:scalarpotential}}
Given the field content of Table~\ref{tab:fields} the most general renormalizable scalar potential compatible with the Pati-Salam symmetry reads
\begin{equation}
    \mathcal{L}_\text{scalar} = V_2+V_3+V_4\;, 
\end{equation}
\begin{widetext}
\begin{align}\label{eq:A1_scalarPotential}
&\begin{aligned}
    V_2 &= m_L^2 \,(\Phi_L)^{\alpha i} \,(\Phi_L^*)_{\alpha i} + m_R^2 \,(\Phi_R)_\alpha^{\;\,a} \,(\Phi_R^*)^{\alpha}_{\;\, a} + \frac{m_\Sigma^2}{2}\, (\Sigma)^{ia}\,(\Sigma^*)_{ia} +\frac{m_\Omega^2}{4} \,(\Omega)^{\alpha \beta}\,(\Omega^*)_{\alpha \beta}\;,\\[1mm]
    V_3 &= \frac{\mu_L}{2}\,\epsilon_{ij}\,(\Phi_L)^{\alpha i}\,(\Phi_L)^{\beta j}\, (\Omega^*)_{\alpha\beta}+\frac{\mu_R}{2}\,\epsilon_{ab}\,(\Phi_R)_\alpha^{\;\, a}\,(\Phi_R)_\beta^{\;\, b}\, (\Omega)^{\alpha\beta} + \mu_{LR}\, (\Phi_L)^{\alpha i}\, (\Phi_R)_\alpha^{\;\, a}\, (\Sigma^*)_{ia}+\hc\;,\\[1mm]
    V_4 &=\lambda_1\,(\Phi_L)^{\alpha i}\,(\Phi_L^*)_{\alpha i}\,(\Phi_L)^{\beta j}\,(\Phi_L^*)_{\beta j}+\lambda_2\,(\Phi_L)^{\alpha i}\,(\Phi_L^*)_{\alpha j}\,(\Phi_L)^{\beta j}\,(\Phi_L^*)_{\beta i}+\lambda_3\,(\Phi_L)^{\alpha i}\,(\Phi_L^*)_{\alpha i}\,(\Phi_R)_\beta^{\;\,a}\,(\Phi_R^*)^\beta_{\;\,a}\\
    &+\lambda_4\,(\Phi_L)^{\alpha i}\,(\Phi_L^*)_{\beta i}\,(\Phi_R)_\alpha^{\;\,a}\,(\Phi_R^*)^\beta_{\;\,a}+\lambda_5\,(\Phi_R)_\alpha^{\;\,a} \,(\Phi_R^*)^{\alpha}_{\;\, a}\,(\Phi_R)_\beta^{\;\,b} \,(\Phi_R^*)^{\beta}_{\;\, b}+\lambda_{6}\,(\Phi_R)_\alpha^{\;\,a} \,(\Phi_R^*)^{\alpha}_{\;\, b}\,(\Phi_R)_\beta^{\;\,b} \,(\Phi_R^*)^{\beta}_{\;\, a}\\
    &+\frac{\lambda_7}{2}\,(\Phi_L)^{\alpha i}\,(\Phi_L^*)_{\alpha i}\,(\Sigma)^{ja}\,(\Sigma^*)_{ja}+\frac{\lambda_{8}}{4}\,(\Phi_L)^{\alpha i}\,(\Phi_L^*)_{\alpha i}\,(\Omega)^{\beta\gamma}\,(\Omega^*)_{\beta\gamma}+\frac{\lambda_{9}}{2}\,(\Phi_R)_\alpha^{\;\,a}\,(\Phi_R^*)^\alpha_{\;\,a}\,(\Sigma)^{jb}\,(\Sigma^*)_{jb}\\
    &+\frac{\lambda_{10}}{4}\,(\Phi_R)_\alpha^{\;\,a}\,(\Phi_R^*)^\alpha_{\;\,a}\,(\Omega)^{\beta\gamma}\,(\Omega^*)_{\beta\gamma}+\frac{\lambda_{11}}{4}\,(\Sigma)^{ia}\,(\Sigma^*)_{ia}\,(\Sigma)^{jb}\,(\Sigma^*)_{jb}+\frac{\lambda_{12}}{8}\,(\Sigma)^{ia}\,(\Sigma^*)_{ia}\,(\Omega)^{\alpha\beta}\,(\Omega^*)_{\alpha\beta}\\
    &+\frac{\lambda_{13}}{16} \,(\Omega)^{\alpha\beta}\,(\Omega^*)_{\alpha\beta}\,(\Omega)^{\gamma\delta}\,(\Omega^*)_{\gamma\delta}+\Bigg[\frac{\lambda_{14}}{2}\,\epsilon_{ij}\,\epsilon_{ab}\,(\Phi_L)^{\alpha i}\,(\Phi_L)^{\beta j}\,(\Phi_R)_\alpha^{\;\, a}\,(\Phi_R)_\beta^{\;\, b}\\
    &+\frac{\lambda_{15}}{4}\,\epsilon_{\alpha\beta\gamma\delta}\,\epsilon_{ij}\,\epsilon^{ab}\,(\Phi_L)^{\alpha i}\,(\Phi_L)^{\beta j}\,(\Phi_R^*)^\gamma_{\;\, a}\,(\Phi_R^*)^\delta_{\;\, b} +\lambda_{16}\,\epsilon_{ij}\,(\Phi_L)^{\alpha i}\,(\Phi_R^*)^\beta_{\;\, a}\, (\Sigma)^{ja}\, (\Omega^*)_{\alpha\beta}+\hc \Bigg]\;.
\end{aligned}
\end{align}
\end{widetext}
where the fields were written in a matrix representation with pairs of indices $(\alpha,\beta), (i,j)$ and $(a,b)$ corresponding to rows and columns within the $SU(4)_C$, $SU(2)_L$ and $SU(2)_R$ fundamental representations, respectively. 

For the fundamental multiplets, the SM components are accommodated within the corresponding Pati-Salam multiplets as
\begin{equation}
\begin{aligned}
    &\!\!(\Phi_L)^{\alpha i} \equiv \left(\begin{array}{c}
         \Upsilon^{xi}  \\
         \Tilde{H'^i} 
    \end{array}\right)\! ,\;(\Phi_R)_{\alpha}^{\;\,a} \equiv \left(\begin{array}{cc}
         T'^*_x & 0  \\
         0 & V_G +S 
    \end{array}\right)\!,\\[3mm] 
    &\qquad\qquad\qquad(\Sigma)^{ia} \equiv \left(\begin{array}{cc}
        H^i & -\Tilde{H}^i \\
    \end{array}\right)\!,\;
        \end{aligned}
\end{equation}
(with the row and column indices corresponding to $SU(4)_C$, and $SU(2)_R$, respectively) and
\begin{equation}
    (\Omega)^{\alpha\beta} \equiv \left(\begin{array}{cc}
       \epsilon^{xyz} \,T_z^*  & T^x \\
        -\,T^y & 0
    \end{array}\right)\,,
\end{equation}
where $(x,y)$ are $SU(3)_c$ row and column triplet indices.

\subsection{Gauge sector}
The $SU(4)_C\times SU(2)_R\to SU(3)_c\times U(1)_Y$ breaking produces nine Goldstone bosons corresponding to the broken generators $T_4^9,\ldots,T_4^{15}$ and $T_R^1,T_R^2,T_R^3$ of $SU(4)_C$ and $SU(2)_R$, respectively. In unitary gauge, these provide nine longitudinal components of the corresponding gauge bosons. Their spectrum is obtained from the broken-phase $\Phi_R$ kinetic term containing
\begin{equation}
\begin{aligned}
    D_\mu\,(\Phi_R)_{\alpha}^{\;\,a} = &\bigg[\partial_\mu \,\delta_\alpha^{\;\,\beta}\,\delta^a_{\;\,b}+i\frac{g_4}{2}\,(\vec{V}\cdot\vec{T_4^*})_{\alpha}^{\;\,\beta}\,\delta^a_{\;\,b}\\
    -&i\frac{g_R}{2}\,\delta_\alpha^{\;\,\beta}\,(\vec{W'}\cdot \vec{T_R})^a_{\;\,b}\bigg](\Phi_R)_{\beta}^{\;\,b}\;;
\end{aligned}
\end{equation}
here the $SU(4)_C$ and $SU(2)_R$ pieces are written in a matrix form to expose the SM-level gauge bosons embedded within:
\begin{equation}
\begin{aligned}
    &\!\!(\vec{V}_\mu\cdot \vec{T}_{4})^{\alpha}_{\;\,\beta} = \frac{1}{2}\left(\begin{array}{cc}
    (\vec{G}\cdot\vec{\lambda})^x_{\;\,y}\! +\! \frac{1}{\sqrt{6}}B'\delta^{xy} & \!\!\!\!\!\! X^x\!\! \\
        X^\dagger_y & \!\!\!\!\!\! -\frac{3}{\sqrt{6}}B'\!\!
    \end{array}\right)_{\!\mu}\!,\\[3mm]
    &\!\!(\vec{W'_\mu}\cdot \vec{T}_{R})^{a}_{\;\,b} = \frac{1}{2}\left(\begin{array}{cc}
        W_0' & W_+' \\
        W_-' & -W_0'
    \end{array}\right)_{\!\mu}\!.
\end{aligned}
\end{equation}
This yields the following expressions for the charged $W'$ and $X$ vector bosons
\begin{equation}
\begin{aligned}
    &M^2_{W'} =\frac{1}{4}g_R^2 V_G^2, \quad M^2_X =\frac{1}{4}g_4^2 V_G^2\;,
\end{aligned}
\end{equation}
along with a mass matrix for $W_0'$ and $B'$:
\begin{equation}
\begin{aligned}
    \mathcal{M}^2_{W_0'B'}=\frac{1}{4}V_G^2\left(\begin{array}{cc}
        g_R^2 & -\sqrt{\frac{3}{2}}\,g_Rg_4 \\
        -\sqrt{\frac{3}{2}}g_Rg_4 & \frac{3}{2}\,g_4^2
    \end{array}\right)\;.
\end{aligned}
\end{equation}
This matrix can be brought into diagonal form by a $2\times 2$ rotation connecting the defining fields to those spanning the mass basis ($Z'$ and $B$) 
\begin{equation}
\begin{aligned}
    \left(\begin{array}{c}
         Z'  \\
         B 
    \end{array}\right) = \left(\begin{array}{cc}
        \cos\theta' & -\sin\theta' \\
        \sin\theta' & \cos\theta'
    \end{array}\right)\left(\begin{array}{c}
         W_0'  \\
         B' 
    \end{array}\right)\;,
\end{aligned}
\end{equation}
\begin{figure}[!htbp]
    \centering
    \includegraphics[scale=1.3]{diagrams/PS_neutrino_mass_two_loop_i.pdf} 
    \includegraphics[scale=1.25]{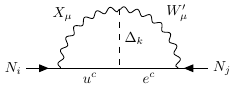} 
    \includegraphics[scale=1.25]{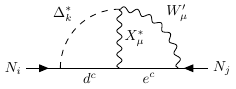}
    \caption{The complete set of two-loop diagrams radiatively generating right-handed neutrino masses in the current model (in unitary-gauge, broken-phase perturbative expansion).}
    \label{fig:neutrinomass2}
\end{figure}
with
\begin{equation}
    \sin\theta' = \frac{g_4}{\sqrt{g_4^2+\frac{2}{3}g_R^2}}\, ,
\end{equation}
leading to a massive $Z'$ boson 
\begin{equation}
M_{Z'}^2 = \frac{1}{4}\left(g_R^2+\frac{3}{2}g_4^2\right)V_G^2\;,\qquad M_{B}^2 = 0\;.
\end{equation}
and a massless $B$ field of the $U(1)_Y$ of the SM, with a gauge coupling 
\begin{equation}
    g_Y = \frac{g_4g_R}{\sqrt{g_4^2+\frac{2}{3}g_R^2}}\;.
\end{equation}
\mbox{}
\subsection{Yukawa sector}
The renormalizable Yukawa Lagrangian of the current model is given by:
\begin{equation}
\begin{aligned}
    \mathcal{L}_Y = &\;Y_{LR} \,\big(\,\overline{\Psi_L}\,\big)_{\alpha i}\,(\Psi_R^c)^{\alpha}_{\;\,a}\,\Sigma^{ia}\\
    + &\; Y_{L}\,\epsilon_{ij} \,\big(\,\overline{\Psi_L^c}\,\big)^{\alpha i}\,(\Psi_L)^{\beta j}\,\Omega^*_{\alpha\beta}\\ 
    +&\; Y_{R}\,\epsilon^{ab} \,\big(\,\overline{\Psi_R}\,\big)^{\alpha}_{\,\;a}\,(\Psi_R^c)^{\beta}_{\,\;b}\,\Omega^*_{\alpha\beta}+\hc\;,
\end{aligned}
\end{equation}
where the matter Pati-Salam multiplets decompose under the SM subgroup into:
\begin{equation}
\begin{aligned}
    &(\Psi_L)^{\alpha i} = \left(\begin{array}{c}
         Q  \\
          L
    \end{array}\right)\;,\;\;(\Psi_R)_{\alpha}^{\;\,a} \equiv \left(\begin{array}{cc}
         d^c & u^c  \\
         e^c & N^c 
    \end{array}\right)\,.
\end{aligned}
\end{equation}
The Majorana mass matrix for the RHNs is radiatively generated at two-loops via diagrams shown in Fig.~\ref{fig:neutrinomass2}.

\subsection{Effective operators}
Among all possible dimension 5 effective operators one can write in the current setting, the following are of our particular interest:
\begin{widetext}
\begin{equation}
\begin{aligned}
    C_{LR1}\,\epsilon^{ij}\,\epsilon^{ab}&\, \big(\,\overline{\Psi_L}\,\big)_{\alpha i} \, \big(\Psi_R^c\big)^\alpha_{\;\, a}\,\big(\Phi_L^*\big)_{\beta j}\, \big(\Phi_R^*\big)^\beta_{\;\,b}\supset V_G\,C_{LR1}\bigg[\bar{L}_i\,e\,H'^{\,i} + (\bar{Q}_i\,d)\,H'^{\,i}\bigg]\;,\\[1mm] 
    C_{LR1}'\,\epsilon^{ij}\,\epsilon^{ab}&\, \big(\,\overline{\Psi_L}\,\big)_{\alpha i} \, \big(\Psi_R^c\big)^\beta_{\;\, a}\,\big(\Phi_L^*\big)_{\beta j}\, \big(\Phi_R^*\big)^\alpha_{\;\,b} \supset V_G\,C_{LR1}'\,\bigg[\bar{L}_i\,e\,H'^{\,i}+\epsilon^{ij}\,\bar{L}_i\, (d\,\Upsilon^*_j)\bigg] \;,\\[1mm] 
    C_{LR2}&\big(\,\overline{\Psi_L}\,\big)_{\alpha i} \, \big(\Psi_R^c\big)^\alpha_{\;\, a}\, \big(\Phi_L\big)^{\beta i}\, \big(\Phi_R\big)_\beta^{\;\,a}\supset V_G\,C_{LR2} \bigg[\,\bar{L}_i\,N\,\Tilde{H'}^{\,i} +(\bar{Q}_i\,u)\,\Tilde{H'}^{\,i} \bigg]\;,\\[1mm] 
   C_{LR2}'\,&\big(\,\overline{\Psi_L}\,\big)_{\alpha i} \, \big(\Psi_R^c\big)^\beta_{\;\, a}\,\big(\Phi_L\big)^{\alpha i}\, \big(\Phi_R\big)_\beta^{\;\,a} \supset  V_G\,C_{LR2}' \,\bigg[\bar{L}_i\,N\,\Tilde{H'}^{\,i}+(\bar{Q}_i\,\Upsilon^i)\,N\bigg]\;,\\[1mm] 
   C_{LL1}&\,\big(\,\overline{\Psi_L^c}\,\big)^{\alpha i} \, \big(\Psi_L\big)^{\beta j}\,\big(\Phi_L^*\big)_{\alpha i}\, \big(\Phi_L^*\big)_{\beta j}\supset  C_{LL1} \,\epsilon_{ij}\epsilon_{kl}\,\bar{(L^i)^c}\,\Tilde{H'}^j\,L^k\,\Tilde{H'}^l\;,\\[1mm] 
   C_{LL1}'&\,\big(\,\overline{\Psi_L^c}\,\big)^{\alpha i} \, \big(\Psi_L\big)^{\beta j}\,\big(\Phi_L^*\big)_{\alpha j}\, \big(\Phi_L^*\big)_{\beta i}\supset C_{LL1}' \,\epsilon_{ij}\epsilon_{kl}\,\bar{(L^i)^c}\,\Tilde{H'}^j\,L^k\,\Tilde{H'}^l\;,\\[1mm] 
   C_{LL2}\,\epsilon_{ij}\,\epsilon_{ab}\,&\big(\,\overline{\Psi_L^c}\,\big)^{\alpha i} \, \big(\Psi_L\big)^{\beta j}\,\big(\Phi_R\big)_{\alpha}^{\;\,a}\, \big(\Phi_R\big)_{\beta}^{\;\,b} = -2\,C_{LL2}\,\epsilon_{ij}\,V_G\,\bar{(L^i)^c}\,(T'^*\,Q^j)\;,\\[1mm] 
    C_{LL3}\,\epsilon_{\alpha\beta\gamma\delta}\,\epsilon_{ij}\,\epsilon^{ab}&\big(\,\overline{\Psi_L^c}\,\big)^{\alpha i} \, \big(\Psi_L\big)^{\beta j}\,\big(\Phi_R^*\big)^{\gamma}_{\;\,a}\, \big(\Phi_R^*\big)^{\delta}_{\;\,b}= 2\, V_G\,C_{LL3} \, \epsilon_{xyz}\,\epsilon_{ij}\,\bar{(Q^{xi})^c}\,Q^{yj}\,T'^z\\[1mm] 
   C_{RR1}\,\epsilon_{\alpha\beta\gamma\delta}\,\epsilon^{ab}\,\epsilon^{cd}\,&\big(\,\overline{\Psi_R}\,\big)^{\alpha}_{\;\,a} \, \big(\Psi_R^c\big)^{\beta}_{\;\,b}\,\big(\Phi_R^*\big)^{\gamma}_{\;\,c}\, \big(\Phi_R^*\big)^{\delta}_{\;\,d} = 4\, V_G\,C_{RR1} \,\epsilon_{xyz}\,\bar{(d^x)^c}\,u^y\,T'^z\,,\\[1mm] 
   C_{RR2}&\,\big(\,\overline{\Psi_R}\,\big)^{\alpha}_{\;\,a} \, \big(\Psi_R^c\big)^{\beta}_{\;\,b}\,\big(\Phi_R\big)_{\alpha}^{\;\,a}\, \big(\Phi_R\big)_{\beta}^{\;\,b}\supset C_{RR2} \bigg[2\,V_G\,\bar{N^c}\,(d\,T'^*) + V_G^2\, \bar{N^c}\,N\bigg]\;,\\[1mm] 
    C_{RR2}'&\,\big(\,\overline{\Psi_R}\,\big)^{\alpha}_{\;\,a} \, \big(\Psi_R^c\big)^{\beta}_{\;\,b}\,\big(\Phi_R\big)_{\alpha}^{\;\,b}\, \big(\Phi_R\big)_{\beta}^{\;\,a}\supset C_{RR2}'\,\bigg[ 2\,V_G\,\bar{e^c}\,(u\,T'^*) + V_G^2\,\bar{N^c}\,N \bigg]\;.
\end{aligned}
\end{equation}    
\end{widetext}
Note that the operators with coefficients $C_{LR1},C_{LR1}',C_{LR2}$ and $C_{LR2}'$ lead to contributions to Yukawa terms for the SM fermions, alongside Dirac mass terms for neutrinos and other BSM cubic operators. Furthermore, those proportional to $C_{LL1}$ and $C_{LL1}'$ lead to Majorana masses for left-handed neutrinos below the electroweak scale via the Weinberg operator, while those parametrized by $C_{RR2}$ and $C_{RR2}'$ contribute to the RHN Majorana masses. Finally, the structures involving $C_{LL2}, C_{LL3}, C_{RR1}, C_{RR2}$ and $C_{RR2}'$ provide additional BSM cubic operators.

\section{Boltzmann equations}\label{app:boltzmann}
To obtain the set of Boltzmann equations corresponding to the scenario described in Sec.~\ref{sec:patisalam} we start from~\cite{Giudice:2003jh}
\begin{equation}
	zHn_{\gamma}\frac{d\eta_X}{dz}=-\sum_{i,j,\dots}[X\cdots\leftrightarrow ij\cdots]\, .
	\label{eq:boltzmann}
\end{equation}
Here $X$ is a particle species, and $z\equiv m_X/T$ for temperature $T$. Furthermore, $\eta_X\equiv n_X/n_\gamma$ is the number density $n_X$ of $X$ normalised to the photon number density
\begin{equation}
    \label{eq:ngamma}
    n_\gamma = \frac{2\zeta(3)}{\pi^2}T^3\, ,
\end{equation}
where $\zeta$ is the Riemann zeta function. Next, 
\begin{equation}
	H =\frac{ 1.66\sqrt{g_*}}{m_{\text{Pl}}}T^2\,, 
\end{equation}
is the Hubble rate in the radiation dominated phase, where $m_\text{Pl}= 1.2\times 10^{19}\text{ GeV}$ is the Planck mass and $g_*$ stands for the number of relativistic degrees of freedom. The RHS of Eq.~\eqref{eq:boltzmann} contains all relevant interaction terms of $X$, where
\begin{equation}
	\label{eq:squarebracket}
    \begin{aligned}
	\![X \cdots \leftrightarrow i\, j \cdots ]&\equiv\frac{\eta_X \cdots}{\eta_X^{\text{eq}} \cdots} \gamma^\eq (X \cdots \to i\, j \cdots)\\
    &-\frac{\eta_i \eta_j \cdots}{\eta_i^{\text{eq}} \eta_j^{\text{eq}} \cdots} \gamma^\eq (i\, j \cdots \to X \cdots)\, .
    \end{aligned}
\end{equation}
Here $i$, $j$, $\cdots$ are particle species that interact with $X$. For relativistic fields we have~\cite{Harvey:1990qw}
\begin{equation}
\label{eq:num2chem}
   n_{i}\approx \frac{g_i T^3}{\pi^2}\times\left\{\begin{array}{ll}
         \displaystyle \zeta(3)+\frac{\mu_i}{T}\zeta(2) & \text{ (boson)}\\[3mm]
         \displaystyle \frac{3}{4}\zeta(3)+\frac{\mu_i}{T}\frac{\zeta(2)}{2} & \text{ (fermion)}
    \end{array}\right.\! ,
\end{equation}
where $\mu_i$ and $g_i$ are the corresponding chemical potential and the number of degrees of freedom, respectively. Note that in the formulae above we have only kept terms up to linear order  in $\mu_i/T$. Next, $\eta^\eq$ therein denotes the relevant equilibrium number density which, for relativistic particles, corresponds to a vanishing chemical potential; for massive particles one has
\begin{equation}\label{eq:etaXeq}
	\eta_X^\eq=\frac{g_X}{4\zeta(3)}z^2K_2\left(z\right)\,.
\end{equation}
Furthermore, $\gamma^\eq$ denotes the equilibrium reaction rate density, which for a decay process is
\begin{equation}\label{eq:gammaeq}
    \gamma^\eq\lr{X \to i\, j\cdots}= n_{X}^{\text{eq}}\frac{K_1\lrs{z}}{K_2\lrs{z}}\Gamma\lr{X \to i\, j\cdots} \, ,
\end{equation}
where $\Gamma\lr{X \to i\, j\cdots}$ is the corresponding partial decay width.
Note that in both \cref{eq:etaXeq,eq:gammaeq}, and also below, $K_n$ denotes the modified Bessel $K$-functions of order $n$.

For two-to-two scattering processes we have
\begin{equation}
\begin{aligned}
	\!\!\gamma^\eq (X\, a \to i\, j) &= \frac{T}{64\pi^4}\\
    &\times\!\!\int_{s_{\min}}^{\infty}\!\!\!\! ds \sqrt{s}\hat{\sigma}(X\, a \to i\, j)K_1\!\!\lr{\frac{\sqrt{s}}{T}},
    \end{aligned}
\end{equation}
where $s_{\min}=\max\big\{{(m_X+m_a)^2,(m_i+m_j)^2}\big\}$, and $\hat{\sigma}(X\, a \to i\, j)$ is the reduced cross section (for details see Appendix~\ref{app:matrixelements}).
Next, we write the total $B-L$ asymmetry density of the SM matter population as
\begin{equation}
\begin{aligned}
    \!n_{B-L}^{\rm SM} \!\equiv\!\sum_i&\frac{1}{3}\big(n_{Q_i}+{n}_{\bar{u^c_i}}+n_{\bar{d^c_i}}\big)-n_{L_i}-{n}_{\bar{e^c_i}}\\
    -&\frac{1}{3}\big(n_{\bar{Q}_i}+{n}_{u^c_i}+n_{d^c_i}\big)+n_{\bar{L}_i}+{n}_{e^c_i}\, .
\end{aligned}
\end{equation}
Note that, though the Weyl spinor $N$ does have a non-zero $B-L$ charge, $n_{B-L}^{\rm SM}$ does not include  $n_N$ since the corresponding Majorana mass term violates $B-L$. Using the chemical potential relations from the SM Yukawa couplings
\begin{equation}
\label{eq:chyuk}
\begin{aligned}
    &\mu_{q_i} = \mu_{\bar{u^c}_i}-\mu_{H_1}\,,\\
    &\mu_{q_i} = \mu_{\bar{d^c}_i}+\mu_{H_1}\,,\\
    &\mu_{L_i} = \mu_{\bar{e^c}_i}+\mu_{H_1}\, , 
\end{aligned}
\end{equation}
the electroweak sphaleron
\begin{equation}
\label{eq:chspah}
    \sum_i\mu_{q_i}=-\frac{1}{3}\sum_i\mu_{L_i}\, ,
\end{equation}
and requiring a total net zero hypercharge
\begin{equation}
    0=\sum_i g_i Y_i\mu_i\, ,
\end{equation}
(where $Y_i$ is the hypercharge of the field $i$ and $g_i$ counts the relevant degrees of freedom), we can relate the chemical potential of each SM particle species to the $\eta_{B-L}^{CM}$ asymmetry as
\begin{align}
    c_{i}\eta_{B-L}^{\rm SM} &\equiv -\frac{\zeta(2)}{\zeta(3)}\frac{\mu_{i}}{T}\, .
\end{align}
Note that we have identified $H_1$ with the SM Higgs $H$.
Explicitly, one then finds
\begin{align}
    c_{H_1} &=\frac{8N}{44N+13}\,,\\
    c_f &= \frac{1}{6}\frac{12N+5}{44N+13}\,,\\
    c_{\bar{d^c}} &= -\frac{12N+1}{44N+13}\,,\\
    c_L &= \frac{12N+3}{44N+13}\, .
\end{align}
where $c_f$ has been defined as
\begin{equation}
    c_f\equiv \frac{1}{6}\lr{c_{\bar{e^c}}-c_{\bar{u^c}}-c_{\bar{d^c}}}\, .
\end{equation}
Finally, we define compound number densities as
\begin{align}
    2\eta_{i^+} &\equiv \eta_{i}+\eta_{i^*}\,,\\
    2\eta_{i^-} &\equiv  \eta_{i}-\eta_{i^*}\,,
\end{align}
where for light fields we have $\eta_{i^+} \to \eta_{i}^\eq$ and $(\eta_{i^-})^2 \approx 0$, which is consistent with keeping up to linear order terms in Eq.~\eqref{eq:num2chem}. For any number of light fields $i$, $j$, $\dots$ we then have
\begin{align}
    \frac{\eta_{i}\eta_{j}\, {\dots} -\eta_{i^*}\eta_{j^*}\, \dots}{\eta_i^\eq\eta_j^\eq\, {\dots}} &\approx 2\frac{\eta_{i^-}}{\eta_i^\eq} + 2\frac{\eta_{j^-}}{\eta_j^\eq}+\, \dots\, ,\\
    \frac{\eta_{i}\eta_{j}\, {\dots}+\eta_{i^*}\eta_{j^*}\, \dots}{\eta_i^\eq\eta_j^\eq\, {\dots}} &\approx 2\, .
\end{align}
Furthermore, for the Higgs field $H_1$ and any SM fermion species $\psi$ we explicitly find
\begin{align}
    \frac{\eta_{H_1^-}}{\eta_{H_1}^\eq} &= -c_{H_1}\eta_{B-L}^{\rm SM}\, ,\\
    \frac{\eta_{\psi^-}}{\eta_{\psi}^\eq} &= -\frac{2}{3}c_{\psi}\eta_{B-L}^{\rm SM}\, ,
\end{align}
respectively. 

\begin{figure}
    \centering
    \raisebox{-0.5\height}{
    \includegraphics[scale=1.25]{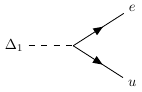}} \hspace{-7mm}
    \raisebox{-0.5\height}{
    \includegraphics[scale=1.25]{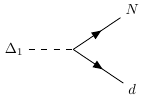}} \hspace{-7mm}
    \raisebox{-0.5\height}{
    \includegraphics[scale=1.25]{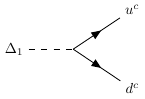}}
    \caption{Decay of $\Delta_1$ into fermions.}
    \label{fig:ddd}
\end{figure}

\begin{figure}
    \centering
    \raisebox{-0.5\height}{
    \includegraphics[scale=1.25]{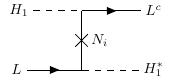}}
    \raisebox{-0.5\height}{
    \includegraphics[scale=1.25]{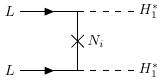}}
    \caption{Washout process mediated by $N$ via $t$-channel.}
    \label{fig:washouttchannel}
\end{figure}

\begin{figure}
    \centering
    \raisebox{-0.5\height}{
    \includegraphics[scale=1.25]{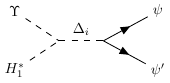}}
    \raisebox{-0.5\height}{
    \includegraphics[scale=1.25]{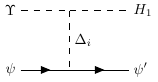}} 
    \caption{Washout processes mediated by $\Delta_i$.}
    \label{fig:deltawashout}
\end{figure}

\begin{figure}
    \centering
    \raisebox{-0.5\height}{
    \includegraphics[scale=1.25]{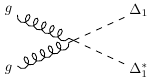}}
    \raisebox{-0.5\height}{
    \includegraphics[scale=1.25]{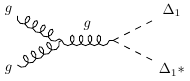}}
    \raisebox{-0.5\height}{
    \includegraphics[scale=1.25]{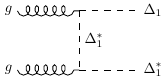}}
    \caption{Gluon fusion into a $\Delta_1$--$\Delta_1^*$ pair.}
    \label{fig:ggfusion}
\end{figure}

For the model described in Sec.~\ref{sec:patisalam}, we find the following relevant reaction rate densities:
\begin{align}
    &\begin{aligned}
    \gamma_D &\equiv  \gamma^\eq\!\lr{\Delta_1\to\Upsilon H_1^*}+\gamma^\eq\!\lr{\Delta_1\to \bar{e^c} \bar{u^c}}\\
    &+\gamma^\eq\!\lr{\Delta_1\to N \bar{d^c}}+\gamma^\eq\!\lr{\Delta_1\to u^c d^c}\,,
    \end{aligned}\\[1mm]
    &\begin{aligned}
    \gamma_{gg} &\equiv  \gamma^\eq\!\lr{\Delta_1\Delta_1^*\to gg}\,,
    \end{aligned}\\[1mm]
    &\begin{aligned}
    \gamma_{\Upsilon D} &\equiv \gamma^\eq\!\lr{\Upsilon\to H_1\bar{e^c}\bar{u^c}}+\gamma^\eq\!\lr{\Upsilon\to H_1 N\bar{d^c}}\\
    &+\gamma^\eq\!\lr{\Upsilon\to H_1 u^cd^c}\,,
    \end{aligned}\\[1mm]
    &\begin{aligned}\label{eq:gammaUpss}
    \gamma_{\Upsilon S} &\equiv  \gamma^\eq_\text{sub}\!\lr{\Upsilon H_1^*\to \bar{e^c}\bar{u^c}}+ \gamma^\eq_\text{sub}\!\lr{\Upsilon H_1^*\to N\bar{d^c}}\\
    &+ \gamma^\eq_\text{sub}\!\lr{\Upsilon H_1^*\to u^cd^c}\,,
    \end{aligned}\\[1mm]
    &\begin{aligned}
    \gamma_{\Upsilon T} &\equiv  \gamma^\eq\!\lr{\Upsilon u^c\to H_1\bar{e^c}}+\gamma^\eq\!\lr{\Upsilon e^c\to H_1\bar{u^c}}\\
    &+\gamma^\eq\!\lr{\Upsilon N^c\to H_1\bar{d^c}}+\gamma^\eq\!\lr{\Upsilon d^c\to H_1 N}\\
    &+\gamma^\eq\!\lr{\Upsilon \bar{u^c}\to H_1 d^c}+\gamma^\eq\!\lr{\Upsilon \bar{d^c}\to H_1 u^c}\,,
    \end{aligned}\\[1mm]
    &\begin{aligned}
    \gamma_{\Upsilon}^{\text{tot}} &\equiv \gamma_{\Upsilon D}+\gamma_{\Upsilon S}+\gamma_{\Upsilon T}\,,
    \end{aligned}\\[1mm]
    &\begin{aligned}
    \gamma_{N_D} &\!\equiv \gamma^\eq\!\lr{N \!\to\! LH_1}\,,
    \end{aligned}\\[1mm]
    &\begin{aligned}
    \gamma_{N_W} &\!\equiv \gamma^\eq\!\lr{LH_1 \!\to\! \bar{L}H_1^*}+ \gamma^\eq\!\lr{LL \!\to\! H_1^*H_1^*}\,,
    \end{aligned}
\end{align}
which, in addition to the diagrams shown in Sec.~\ref{sec:patisalam}, correspond to the diagrams shown in~\cref{fig:ddd,fig:washouttchannel,fig:deltawashout,fig:ggfusion}. 
Note that we have neglected two-to-two washout processes with external $\Delta_1$. For the scalar decay modes we then have
\begin{align}
    &\begin{aligned}
    (r+\epsilon)\gamma_D &= \gamma^\eq\!\lr{\Delta_1\to\Upsilon H_1^*}\,,
    \end{aligned}\\[1mm]
    &\begin{aligned}
    (r-\epsilon)\gamma_D &= \gamma^\eq\!\lr{\Delta_1^*\to\Upsilon^* H_1}\,,
    \end{aligned}
\end{align}
and similarly for the fermion modes
\begin{eqnarray}
    (1-r-\epsilon)\gamma_D &=& \gamma^\eq\!\lr{\Delta_1\to \bar{e^c} \bar{u^c}}
    +\gamma^\eq\!\lr{\Delta_1\to N \bar{d^c}}\nonumber\\
    &+&\gamma^\eq\!\lr{\Delta_1\to u^c d^c}\,,\\
     (1-r+\epsilon)\gamma_D &=& \gamma^\eq\!\lr{\Delta_1^*\to e^c u^c}+\gamma^\eq\!\lr{\Delta_1^*\to N^c d^c}\nonumber\\
    &+&\gamma^\eq\!\lr{\Delta_1^*\to \bar{u^c} \bar{d^c}}\,.
\end{eqnarray}
Furthermore, in Eq.~\eqref{eq:gammaUpss}, the \textit{sub} symbol denotes on-shell subtraction, such that
\begin{equation}
    \begin{aligned}
    \gamma^\eq_\text{sub}\lr{\Upsilon H_1^* \to \psi\psi'} &\equiv \gamma^\eq\lr{\Upsilon H_1^* \to \psi\psi'}\\
    &-\gamma^\eq_\text{on-shell}\lr{\Upsilon H_1^* \to \psi\psi'}\, ,
    \end{aligned}
\end{equation}
where $\psi\psi'\in\{d^cu^c,\, N d,\, e u\}$. Using (cf. Ref.~\cite{Fridell:2021gag})
\begin{equation}
    \gamma^\eq_\text{on-shell}\lr{\Upsilon H_1^* \to \psi\psi'}\approx\frac{1}{4}\gamma_D\lr{r-r^2-\epsilon}\,,
\end{equation}
we then have
\begin{align}
    \gamma^\eq_\text{sub}\!\lr{\Upsilon H_1^* \to \psi\psi'}\!=&
    \frac{1}{4}\lr{\gamma_{\Upsilon S}-\gamma_D(r-r^2-\epsilon)}\,,\\
    \gamma^\eq_\text{sub}\!\lr{\Upsilon^* H_1 \to \psi\psi'}\!=&
    \frac{1}{4}\lr{\gamma_{\Upsilon S}-\gamma_D(r-r^2+\epsilon)}\! .
\end{align}
Note that for $d^cu^c$ we have an extra factor 2 due to color combinatorics. 

We then find the following set of Boltzmann equations:
\begin{widetext}
    \begin{align}
    &\begin{aligned}\label{eq:Dpeq}
    zHn_\gamma \frac{d\eta_{\Delta_1^+}}{dz} = &-\gamma_D\Bigg[\frac{\eta_{\Delta_1^+}}{\eta^\eq_{\Delta_1}}-1
    + \epsilon\lr{\frac{\eta_{\Upsilon^-}}{\eta_{\Upsilon^-}^\eq}+c_{H_1}+c_f+\frac{c_{\bar{d^c}}}{6}\lr{\frac{\eta_N}{\eta_N^\eq}-1}}\eta_{B-L}^{\rm SM}\Bigg]-\gamma_{gg}\Bigg[\lr{\frac{\eta_{\Delta_1^+}}{\eta_{\Delta_1}^\eq}}^2-1\Bigg]\,,
    \end{aligned}\\
    &\begin{aligned}\label{eq:Dmeq}
    zHn_\gamma \frac{d\eta_{\Delta_1^-}}{dz} = &-\gamma_D\Bigg[\frac{\eta_{\Delta_1^-}}{\eta^\eq_{\Delta_1}}
    -r\frac{\eta_{\Upsilon^-}}{\eta_{\Upsilon}^\eq}-\lr{rc_{H_1}-(1-r)\lr{c_f+\frac{c_{\bar{d^c}}}{6}\lr{\frac{\eta_N}{\eta_N^\eq}-1}}}\eta_{B-L}^{\rm SM}\Bigg]\,,
    \end{aligned}\\
    &\begin{aligned}\label{eq:Yeq}
    zHn_\gamma \frac{d\eta_{\Upsilon^-}}{dz} = &-\gamma_D\epsilon\Bigg[\frac{\eta_{\Delta_1^+}}{\eta^\eq_{\Delta_1}}-1
    \Bigg]-\lr{r^2\gamma_D+\gamma_{\Upsilon}^\text{tot}}\frac{\eta_{\Upsilon^-}}{\eta_{\Upsilon}^\eq}+\gamma_Dr\frac{\eta_{\Delta_1^-}}{\eta_{\Delta_1}^\eq}\\
    &-\Bigg[\big(rc_{H_1}-(1-r)c_f\big)r\gamma_D+(c_{H_1}+c_f)\gamma_{\Upsilon}^\text{tot}-\lr{(1-r)r\gamma_D-\gamma_{\Upsilon}^\text{tot}}\frac{c_{\bar{d^c}}}{6}\lr{\frac{\eta_N}{\eta_N^\eq}-1}\Bigg]\eta_{B-L}^{\rm SM}\,,
    \end{aligned}
\end{align}
\begin{align}
    &\begin{aligned}
    zHn_\gamma \frac{d\eta_{N}}{dz} = &-\frac{1}{4}\Bigg[(1-r)^2\gamma_D+4\gamma_{N_D}+\gamma_{\Upsilon}^\text{tot}\Bigg]\lr{\frac{\eta_N}{\eta_N^\eq}-1}+\frac{1}{4}(1-r)\gamma_D\lr{\frac{\eta_{\Delta_1^+}}{\eta^\eq_{\Delta_1}}-1}\\
    &-\frac{\epsilon}{4}\gamma_D\lr{\frac{\eta_{\Upsilon^-}}{\eta_{\Upsilon}^\eq}+c_{H_1}\eta_{B-L}^{\rm SM}-\frac{\eta_{\Delta_1^-}}{\eta^\eq_{\Delta_1}}}\,,
    \end{aligned}\\
    &\begin{aligned}\label{eq:BLeq}
    6zHn_{\gamma}\frac{d\eta_{B-L}^{\rm SM}}{dz} = &-5\gamma_D\epsilon\lr{\frac{\eta_{\Delta_1^+}}{\eta_{\Delta_1}^\eq}-1}+5\gamma_D(1-r)\frac{\eta_{\Delta_1^-}}{\eta_{\Delta_1}^\eq}+5\Big(r(1-r)\gamma_D-\gamma_{\Upsilon}^\text{tot}\Big)\frac{\eta_{\Upsilon^-}}{\eta_{\Upsilon}^\eq}\\
    &-\Bigg[(1-r)\lr{5rc_{H_1}+(1-r)\lr{8c_f-\frac{4}{3}c_{\bar{d^c}}}}\gamma_D+\lr{8c_f-\frac{4}{3}c_{\bar{d^c}}+5c_{H_1}}\gamma_{\Upsilon}^\text{tot}\\
    &-\frac{2}{3}c_{\bar{d^c}}\Big((1-r)^2\gamma_D+\gamma_{\Upsilon}^\text{tot}\Big)\frac{\eta_N}{\eta_N^\eq}+(12c_{H_1}+8c_L)\gamma_{N_W}\Bigg]\eta_{B-L}^{\rm SM}\, .
    \end{aligned}
    \end{align}
\end{widetext}
Here we see that $\eta_{\Delta_1^-}$ rapidly decays when $\gamma_D$ is strong (similar to $\eta_{\Delta_1^+}$), while the generation of its abundance only arises as a secondary effect via $\eta_{\Upsilon^-}$ and $\eta_{B-L}$. We therefore expect $\eta_{\Delta_1^-}$ to be small throughout its evolution and to not have a significant impact on the other particle densities. Furthermore, $\eta_N$ only enters the equations for $\eta_{\Upsilon^-}$ and $\eta_{B-L}^\text{SM}$ while coupled to $\eta_{B-L}^\text{SM}$. Since $\eta_{B-L}^\text{SM}$ is generally small in the regions of parameter space that we are ultimately interested in (in order to match with the observed value), we do not expect $\eta_N$ to have a significant impact either.

Note that the effective factor $5/6$ in Eq.~\eqref{eq:BLeq} differs from the expected $4/3$ (c.f.\ Sec.~\ref{sec:baryogenesis}) since we have not counted the $B-L$ charge carried by $N$ in our definition of $n_{B-L}^{\rm SM}$. Since the same Pati-Salam-level Yukawa coupling leads to all the fermionic decay modes of $\Delta_1$, we can relate their branching ratios as
\begin{equation}
\begin{aligned}
\label{eq:BReq}
    \text{BR}(\Delta_1\to u e) &=\text{BR}(\Delta_1\to d N)\\
    &=\tfrac{1}{2}\text{BR}(\Delta_1\to u^c d^c),
\end{aligned}
\end{equation}
where the factor $\frac{1}{2}$ is due to color combinatorics, and where we have taken the massless limit for $N$. Counting the $B-L$ charges we see that each fermionic final state in $\Delta_1$-decays has $B-L=-\frac{2}{3}$, leading to a factor $\frac{4}{3}$ when comparing it to the $\Upsilon^-$ number density (due to the factor 2 in the definition of $\eta_{\Upsilon^-}$). However, not counting the $B-L$ charge of $N$ gives us $B-L=-\frac{3}{4}\times\frac{2}{3}+\frac{1}{4}\times\frac{1}{3}=-\frac{5}{12}$, where the $dN$ state (which appears in $\frac{1}{4}$ of the fermionic $\Delta_1$ decays) now has $B-L=\frac{1}{3}$, leading finally to a factor $\frac{5}{6}$ relative to $\Upsilon^-$. Note further that in \cref{eq:Dpeq,eq:Dmeq,eq:Yeq} we have neglected terms proportional to $(\eta_{N}+\eta_{N^c})$ arising from on-shell subtraction, as well as a subtle difference in the $s$- vs.\ $t$-channel contributions to $\gamma^\text{tot}_\Upsilon$ in Eq.~\eqref{eq:Yeq}.

\section{Matrix elements}\label{app:matrixelements}
Here we list the matrix elements for the processes relevant to the baryogenesis mechanism in Sec.~\ref{sec:baryogenesis}. For the decays we have 
\begin{align}   
    &\Gamma(A\to bc) = \frac{\lambda^{1/2}(M_A^2,m_b^2,m_c^2)}{16\pi M_A^3}\Big|\mathcal{M}(A\to bc)\Big|^2 ,\\
    &\begin{aligned}
    \Gamma(\Upsilon\!\to \psi \psi' H_1) &= \frac{1}{(2\pi)^3}\frac{1}{32 m_\Upsilon^3}\\
    &\times\!\!  \int\!\! |\mathcal{M}(\Upsilon\!\to \psi \psi' H_1)|^2 dm_{12}^2 dm_{13}^2 ,
    \end{aligned}
\end{align}
where $(m_{12}^2)_\text{min} = 0$, $(m_{12}^2)_\text{max} = m_\Upsilon^2$, $(m_{13}^2)_\text{min} = 0$ and $(m_{13}^2)_\text{max} = m_\Upsilon^2-m_{12}^2$. The matrix elements are
\begin{align}
    &|\mathcal{M}(\Delta_1\to \Upsilon H_1^*)|^2 = 3\mu_{LR}^2 \,|U_{\Delta12}|^2\,|U_{H11}|^2\,, \\[1mm]
    &|\mathcal{M}(\Delta_1\to \psi \psi')|^2 =12 |Y_R|^2|U_{\Delta 11}|^2m_{\Delta_1}^2\,,  \\[1mm]
    &|\mathcal{M}(N_1\to L H_1)|^2 =2|Y_{LR}|^2|U_{H11}|^2 m_{N_1}^2\,, \\[1mm]
    &\begin{aligned}
    |\mathcal{M}(\Upsilon\to  \psi \psi' H_1)|^2 &= 12\mu_{LR}^2|U_{H11}|^2|Y_R|^2\\
    &\times\,\frac{|U_{\Delta11}U_{\Delta12}^*|^2}{m_{\Delta 1}^4}m_{12}^2 \, ,
    \end{aligned}
\end{align}
where $\psi\psi'\in\{d^cu^c,\, N d,\, e u\}$. 
For two-to-two scattering we have
\begin{equation}
\begin{aligned}
    &\hat\sigma(A B\to C D)=\frac{1}{8\pi }\int_{t^-}^{t^+}|\mathcal{M}(A B\to C D)|^2 dt,
\end{aligned}
\end{equation}
with the integration limits
\begin{equation}
	\begin{aligned}
		t^\pm &= \frac{1}{4s}(m_A^2-m_B^2-m_C^2+m_D^2)^2\\
        &-\frac{1}{4s}\lr{\lambda^{1/2}(s,m_A^2,m_B^2)\mp\lambda^{1/2}(s,m_C^2,m_D^2)}^2\! ,
	\end{aligned}
\end{equation}
where $\lambda(x,y,z)=x^2+y^2+z^2-2xy-2xz-2yz$ is the Källén function. The matrix elements for the $\Delta_1$-mediated washouts are given by
\begin{align}
     &\begin{aligned}
     |\mathcal{M}(\Upsilon H_1^* \to  \psi\psi')|^2 &= 12\mu_{LR}^2|U_{H11}|^2|Y_R|^2\\
     &\times \frac{|U_{\Delta11}U_{\Delta12}^*|^2}{(s-m_{\Delta_1}^2)^2}s,
     \end{aligned}\\
     &\begin{aligned}
     |\mathcal{M}(\Upsilon \psi^c\to H_1 \psi')|^2 &= -12\mu_{LR}^2|U_{H11}|^2|Y_R|^2\\
     &\times \frac{|U_{\Delta11}U_{\Delta12}^*|^2}{(t-m_{\Delta 1}^2)^2}t\,,
     \end{aligned}
\end{align}
with similar contributions from $\Delta_2$, and
\begin{align}
     &\begin{aligned}
     |\mathcal{M}(H_1 L\to H_1^* L^c)|^2 &= 4|U_{H11}|^4|Y_{LR}^\dagger Y_{LR}|^2\\
     &\times  \frac{m_{N}^2s}{(s-m_{N}^2)^2}, 
     \end{aligned}\\
     &\begin{aligned}
     |\mathcal{M}(L L\to H_1^* H_1^*)|^2 &= -4\,|U_{H11}|^4|Y_{LR}^\dagger Y_{LR}|^2\\
     &\times  \frac{m_{N}^2t}{(t-m_{N}^2)^2}\, ,
\end{aligned}
\end{align}
for the washouts mediated by $N$. 
Finally, for the two-body decay of $\Upsilon$ in Fig~\ref{fig:Upsilondecay} (left), and for the gluon annihilation shown in Fig~\ref{fig:ggfusion}, we find, respectively, 
\begin{widetext}
    \begin{align}\label{eq:UpsilontoSMGamma}
        \Gamma\lr{\Upsilon\to d L^c } &=  \sum_{ij}\frac{1}{512}\frac{1}{(2\pi)^5}\frac{|\mu_{LR}|^2m_\Upsilon^2 m_{N_j}}{(m_{\Delta_i}^2-m_{N_j}^2)^2}|(U_{\Delta})_{i2}|^2\log\lr{\frac{m_{\Delta_i}^2}{m_{N_j}^2}}\lr{2Y_R^\dagger Y_{LR}}_{jj}^2\, ,\\
        \hat{\sigma}\lr{\Delta_1\Delta_1^*\to gg}&=\frac{g_s^4}{384 \pi  s^2} \left(\!\big(62 m_{\Delta_1}^2+5s\big) \sqrt{s^3 \left(s-4 m_{\Delta_1}^2\right)}-16 m_{\Delta_1}^2 \left(m_{\Delta_1}^2+4 s\right) \tanh
   ^{-1}\!\!\left(\sqrt{1-\frac{4 m_{\Delta_1}^2}{s}}\right)\right) . 
    \end{align}
\end{widetext}

\bibliographystyle{apsrev4-1}
\bibliography{references}

\end{document}